\documentclass[aps,rmp,reprint,showpacs,showkeys,superscriptaddress,amsmath,amssymb,floatfix,author-numerical,nofootinbib]{revtex4-1}
\def\MyClass{0} % this is during the typesetting of the document
\def\MyFigSize{0.49\textwidth}
\def\MyFigSizeStar{0.99\textwidth}
\ifnum\MyClass=0
\usepackage[T1]{fontenc}
\usepackage[utf8]{inputenc}
\usepackage[english]{babel}
\usepackage{times}
\usepackage{amsmath,amsfonts,amssymb} % AMS
\usepackage[pdftex]{graphicx} %we use PDF figures
\usepackage[usenames,dvipsnames,svgnames,table,rgb]{xcolor}
\fi
\ifnum\MyClass=1
\usepackage[T1]{fontenc}
\usepackage[utf8]{inputenc}
\usepackage[english]{babel}
\usepackage{times}
\usepackage{amsmath,amsfonts,amssymb} % AMS
\usepackage[pdftex]{graphicx} %we use PDF figures
\usepackage[usenames,dvipsnames,svgnames,table,rgb]{xcolor}
\fi

\newcommand{\vect}[1]{\boldsymbol{#1}} % for vectors
\newcommand{\avg}[1]{\left\langle #1 \right\rangle} % for average
\newcommand{\ket}[1]{\left| #1 \right\rangle} % for Dirac kets
\newcommand{\bra}[1]{\left\langle #1 \right|} % for Dirac bras
\newcommand{\oin}{\omega_{\rm in}} % for incoming radiation
\newcommand{\oout}{\omega_{\rm out}} % for outgoing radiation
\newcommand{\cpolr}{\textcolor{red}{$\circlearrowright$}} % right circular polarization
\newcommand{\cpoll}{\textcolor{blue}{$\circlearrowleft$}} % left circular polarization
\newcommand{\ctc}[1]{CTC-#1} % CORE-TO-CORE
\newcommand{\vtc}[1]{VTC-#1} % VALENCE-TO-CORE
\newcommand{\rp}{$\frac{E}{\Delta E}$} %resolving power
\newcommand{\ts}[3]{$^{\rm #1}\mathrm{#2}_{\rm #3}$} % TERM SYMBOL
\newcommand{\dms}[3]{#1$_{1-x}$#2$_{x}$#3} % TERNARY ALLOY AS DMS
\newcommand{\ddms}[4]{#1$_{1-x-y}$#2$_{x}$#3$_{y}$#4} % QUATERNARY ALLOY AS DOPED-DMS
\newcommand{\eg}{e.g.\ }
\newcommand{\ie}{i.e.\ }
\newcommand{\cf}{cf.\ }
\newcommand{\vs}{vs\ }
\ifnum\MyClass=1
\newcommand{\onlinecite}[1]{\cite{#1}} 
\fi
\ifnum\MyClass=2
\newcommand{\onlinecite}[1]{\cite{#1}} 
\fi
\usepackage{hyperref}
\hypersetup{%
  colorlinks=true,
  linkcolor=blue,
  urlcolor=blue,
  citecolor=blue,
  pdftitle={Hard x-ray emission spectroscopy: a powerful tool for the characterization of magnetic semiconductors},%
  pdfauthor={M. Rovezzi and P. Glatzel},%
  pdfsubject={spintronics, magnetic semiconductors synchrotron characterization, x-ray emission spectroscopy, resonant inelastic x-ray scattering},%
  pdfkeywords={spintronics, magnetic semiconductors, synchrotron characterization, x-ray emission spectroscopy, resonant inelastic x-ray scattering}%
}
\ifnum\MyClass=2
\JournalInfo{Semicond. Sci. Technol. 29 (2014) 023002} % Journal information
\Archive{Author's version, \href{http://dx.doi.org/10.1088/0268-1242/29/2/023002}{DOI:10.1088/0268-1242/29/2/023002}} % Additional notes (e.g. copyright, DOI, review/research article)

\PaperTitle{Hard x-ray emission spectroscopy: a powerful tool for the characterization of magnetic semiconductors} % Article title

\Authors{M Rovezzi\textsuperscript{1}*, P Glatzel\textsuperscript{1}} % Authors
\affiliation{\textsuperscript{1}\textit{ European Synchrotron Radiation Facility, 6 rue Jules Horowitz, 38043 Grenoble, France}} % Author affiliation
\affiliation{*\textbf{Corresponding author}: \href{mailto:mauro.rovezzi@esrf.eu}{mauro.rovezzi@esrf.eu}} % Corresponding author

 % Keywords - if you don't want any simply remove all the text between the curly brackets
\newcommand{\keywordname}{Keywords} % Defines the keywords heading name
\fi
\begin{document}
\ifnum\MyClass=0
\title[]{Hard x-ray emission spectroscopy: a powerful tool for the characterization of magnetic semiconductors}
\author{M.~Rovezzi}\email{mauro.rovezzi@esrf.eu}\homepage[this is an author-created, un-copyedited version of an article accepted for publication in Semiconductor Science and Technology. IOP Publishing Ltd is not responsible for any errors or omissions in this version of the manuscript or any version derived from it. The definitive publisher-authenticated version is available online at \href{http://dx.doi.org/10.1088/0268-1242/29/2/023002}{doi:10.1088/0268-1242/29/2/023002}; please cite this paper as: {\sf Semicond. Sci. Technol. 29 (2014) 023002}.]{}
\affiliation{European Synchrotron Radiation Facility, 6 rue Jules Horowitz, 38043 Grenoble, France}
\author{P.~Glatzel}%\email{pieter.glatzel@esrf.eu}
\affiliation{European Synchrotron Radiation Facility, 6 rue Jules Horowitz, 38043 Grenoble, France}
\date{\today}
\fi
% </revtex4-1>
%
% <iopart>
\ifnum\MyClass=1
\topical[]{Hard x-ray emission spectroscopy: a powerful tool for the characterization of magnetic semiconductors}
\author{M. Rovezzi$^1$ and P. Glatzel$^1$}
\address{$^1$ European Synchrotron Radiation Facility, 6 rue Jules Horowitz, 38043 Grenoble, France}
\eads{\mailto{mauro.rovezzi@esrf.eu}}%, \mailto{pieter.glatzel@esrf.eu}}
\date{\today}
\submitto{\SST}
\fi
% </iopart>
%
% <myiopart>
\ifnum\MyClass=0
\begin{abstract}
\fi
\ifnum\MyClass=1
\begin{abstract}
\fi
\ifnum\MyClass=2
\Abstract{
\fi
  This review aims to introduce the x-ray emission spectroscopy (XES)
  and resonant inelastic x-ray scattering (RIXS) techniques to the
  materials scientist working with magnetic semiconductors (\eg
  semiconductors doped with 3$d$ transition metals) for applications
  in the field of spin-electronics. We focus our attention on the hard
  part of the x-ray spectrum (above 3 keV) in order to demonstrate a
  powerful element- and orbital-selective characterization tool in the
  study of bulk electronic structure. XES and RIXS are
  photon-in/photon-out second order optical processes described by the
  Kramers-Heisenberg formula. Nowadays, the availability of third
  generation synchrotron radiation sources permits applying such
  techniques also to dilute materials, opening the way for a detailed
  atomic characterization of impurity-driven materials. We present the
  K$\beta$ XES as a tool to study the occupied valence states
  (directly, via valence-to-core transitions) and to probe the local
  spin angular momentum (indirectly, via intra-atomic exchange
  interaction). The spin sensitivity is employed, in turn, to study
  the spin-polarised unoccupied states. Finally, the combination of
  RIXS with magnetic circular dichroism (RIXS-MCD) extends the
  possibilities of standard magnetic characterization tools.\\
%
%% \\
%
%% (Some figures may appear in colour only in the online journal)
%
\ifnum\MyClass=0
\end{abstract}
\fi
\ifnum\MyClass=1
\end{abstract}
\fi
\ifnum\MyClass=2
}
\fi

%<revtex4-1>
\ifnum\MyClass=0
\maketitle
\fi
%</revtex4-1>
% <myiopart>
\ifnum\MyClass=2
\flushbottom % Makes all text pages the same height
\maketitle % Print the title and abstract box
%%\tableofcontents % Print the contents section
\thispagestyle{empty} % Removes page numbering from the first page
\fi
% </myiopart>

%
%\tableofcontents
%
%========================================%
\section{Introduction}
\label{sec:intro}
%========================================%
% semiconductor spintronics
Semiconductors doped with few percent (10$^{20}$--10$^{21}$ at/cm$^3$)
of magnetic elements such as transition metals (TM) or rare earth
elements (RE) are promising building blocks for semiconductor-based
spin-electronics \cite{Dietl:2010_NMat,Dietl:2013_arxiv} ({\em
  spintronics}). In the dilute magnetic semiconductor (DMS) model, the
TM (RE) dopants randomly substitute in the host semiconductor and, due
to the unpaired $d$ ($f$) states, bring a local net magnetic
moment. These local moments, via inter-atomic exchange interactions
(eventually mediated by defects or carriers), bring magnetic
properties to the semiconductor, leading to an overall half-metallic
behavior \cite{Coey:2004_JPDAP}, that is, the presence of spin
polarization at the Fermi level. Such materials can be used then as
injector or detector for spin-polarised currents in semiconductors and
permit realising spintronics devices as, for example, the proposed
spin field-effect transistor \cite{Datta:1990_APL}, overcoming the
conductivity mismatch problem \cite{Schmidt:2000_PRB} that arises for
ferromagnetic-metal/semiconductor hetero-structures. It is important
to clarify that non-magnetic semiconductors such as II-VI or III-V
alloys (\eg GaAs, GaN or ZnO) have been identified historically as
host materials for DMS, because their epitaxial growth is of high
quality and they can be easily integrated in CMOS technology (what is
currently used for constructing integrated circuits). Recently, pushed
by the advances in epitaxial growth of oxide materials
\cite{Opel:2012_JPDAP,Bibes:2011_AIP}, also bulk magnetic oxides such
as transition metal oxides are considered for semiconductor
spintronics. We will focus mainly on DMS because these materials
represent an ideal workbench for testing new and exciting effects as
quantum spintronics
\cite{Awschalom:2013_S,Koenraad:2011_NatMater,George:2013_PRL} (also
known as {\em solotronics}) or the spin solar cell
\cite{Jansen:2013_NatMater,Endres:2013_NatComms} and others
\cite{Sinova:2012_NatMater}.\\
% DMS vs CMS
DMS currently suffer from low (ferro)magnetic transition
temperatures. In order to obtain the magnetic coupling persisting well
above room temperature, the concentration of the active dopants is
pushed (in most cases) far above the thermodynamic solubility limit by
out-of-equilibrium epitaxial growth methods (\eg low temperature
molecular beam epitaxy). This can cause side effects such as the
incorporation of counter-productive defects (\eg Mn interstitials in
\dms{Ga}{Mn}{As}) or a chemical phase separation, where the density of
the magnetic impurities is not constant over the host crystal
(condensed magnetic semiconductors, CMS). Two recent reviews
\cite{Bonanni:2010_CSR,Sato:2010_RMP} describe in detail the status
of current research on DMS/CMS both from the experimental and
theoretical point of view. They show a growing consensus that
theorical results can drive the experiments in the optimization of new
and exciting materials only if an accurate characterization at the
nano-scale and at bulk level is put in place
\cite{Zunger:2010_Phys}.\\
% intro XES/RIXS toolkit in DMS/CMS
In order to tackle this point, we review a spectroscopic technique,
the hard x-ray emission spectroscopy with synchrotron radiation
(resonant inelastic x-ray scattering, RIXS) that is a powerful tool in
characterizing such materials. It is a direct feedback for the
scientists who need to engineer their materials at the nano-scale
(bottom-up approach) via a fine control of their atomic and electronic
structure. This permits realizing relevant devices and to explore new
ideas and concepts in spintronics. The application of RIXS to doped
semiconductors is stimulating also for the theoreticians aiming to
calculate experimental (spectroscopic) observables. In fact, RIXS
permits combining two theoretical approaches to the description of the
electronic structure of matter: band calculations based on the density
functional theory (DFT) and atomic calculations based on the ligand
field multiplet theory (LFMT). In fact, on one hand, DMS are well
described by DFT as periodic systems and, on the other hand, DMS can
be modeled by the LFMT model as a deep impurity in a crystal field.\\
% XAS intro
Being naturally element and spin/orbital angular momentum selective,
x-ray spectroscopy permits studying the source of the observed
macroscopic magnetism from a local structural and electronic point of
view. X-ray absorption spectroscopy (XAS) is one of the well
established and widely used tools in x-ray spectroscopy. In XAS, an
incoming photon (of energy $\hbar\oin$) excites an inner-shell
electron to an unoccupied level, leaving the system in an excited
state with a core hole that lives for a certain time, $\tau$, that is
linked to the uncertainty in its energy, $\Gamma$, via the Heisenberg
principle: $\Gamma\tau\ge\hbar/2$ (\eg a lifetime of 1 fs implies a
broadening of $\approx$ 0.1 eV). Experimentally, XAS is observed as
discontinuities (the absorption edges) in the absorption coefficient,
$\mu(\hbar\oin)$. In a one-electron picture, the absorption edges
mainly arise from electric dipole transitions ($\Delta l$~=~$\pm 1$),
that is, transitions to the empty partial density of states (PDOS) -
the density of states projected on the orbital angular momentum, $l$,
of the absorbing atom. Thus, the orbitals with $p$ symmetry are probed
in K, L$_1$ and M$_1$ edges ($s \to p$), the $d$ in the L$_{2,3}$ and
M$_{2,3}$ ($p \to d$) and the $f$ in the M$_{4,5}$ ($d \to f$). By
scanning the incoming energy around the absorption edge of a given
element in the sample, the spectroscopist can describe the atomic and
electronic structure of the system, either via a fingerprint approach,
based on the use of model compounds, or supported by calculations,
based on quantum mechanics. The emitted photoelectron wave can be
viewed as scattering with the neighboring atoms and interfering with
itself. This gives rise to the fine structure observed in the
absorption coefficient. The XANES (x-ray absorption near-edge
structure) and EXAFS (extended x-ray absorption fine structure)
techniques \cite{Lee:1981_RMP}, described by the multiple scattering
theory \cite{Rehr:2000_RMP}, permit extracting the local
geometry/symmetry and the bond distances, plus the coordination
numbers and disorder from the analysis of the fine structure. XANES
and EXAFS have been successfully applied to the geometric structure
analysis in semiconductor heterostructures
\cite{Boscherini:2008_book}, DMS/CMS
\cite{Rovezzi:2009_thes,DAcapito:2011_SST} and low-dimensional systems
\cite{Mino:2013_JPDAP}.\\
% XAS -> XMCD -> soft vs hard x-rays
XAS can be used also as an element-selective magnetometer by recording
the difference in absorption of linearly/circularly polarised light in
a presence of a magnetic field, the x-ray magnetic linear/circular
dichroism (XMLD/XMCD) technique \cite{Stohr:1999_JMMM}. This is an
advantage with respect to those techniques where the whole sample
response to an external perturbation is measured (\eg superconducting
quantum interference device magnetometry \cite{Sawicki:2011_SST} or
electron paramagnetic resonance \cite{Wilamowski:2011_SST}). With
respect to DMS/CMS, XMCD was successfully combined with the x-ray
(natural) linear dichroism \cite{Brouder:1990_JPCM} (XLD) and
systematically applied to the study of \dms{Zn}{Co}{O} to link the
local magnetic and structural properties \cite{Ney:2010_NJP}. For 3$d$
TM, XMCD is usually measured at the L edges, residing in the soft
x-ray region (below 1 keV) or at the K edges, residing in the hard
part of the spectrum (above 3 keV). XMCD at the L edge has the
advantage of accessing the partially filled $d$ orbitals via direct
electric dipole transitions and the possibility to separate the spin
and orbital contribution to the magnetic moment via sum rules
\cite{Thole:1992_PRL,Carra:1993_PRL}. XMCD at the K edge probes only
the orbital component and results in a very small signal ($\approx
10^{-3}$ times smaller that XLD). The advantages in using hard x-rays
consist in the sample environment and the bulk sensitivity. A vacuum
environment around the sample is not required with hard x-rays, thus
it is possible to measure {\em in operando} devices or in extreme
conditions (\eg high pressure). Furthermore, the higher penetration
depth permits probing bulk properties and access buried interfaces or
superstructures (\eg two-dimensional electron/hole gases) that are the
relevant structures of real devices to study spin transport
mechanisms. Soft x-rays are suitable in the case of thin films (few
tens of nm thick) deposited on a substrate, where the electron yield
(EY) detection is used as surface probe, while the fluorescence yield
(FY) as representative of the full thickness. Nevertheless, FY suffers
from strong self-absorption effects and is not a true measurement of
the linear absorption coefficient as obtained in transmission
measurements or EY
\cite{DeGroot:1994_SSC,Kurian:2012_JPCM}. This has relevant
consequences on the study of magnetic materials with soft x-rays XMCD
because it means that it is not possible to compare EY measurements to
FY ones and, most importantly, it implies the non-applicability of sum
rules. An alternative method based on x-ray emission has been proposed
recently \cite{Achkar:2011_PRB,Achkar:2011_SciRep}. To overcome those
difficulties, the use of a hard x-ray probe in an inelastic scattering
configuration is gaining momentum. By working in an energy loss scheme
(inelastic scattering) it is possible to reach the same final states
reachable with soft x-rays in a second order process, that is, by
passing via an intermediate state that is excited resonantly, strongly
enhancing the spectral features \cite{Carra:1995_PRL}.\\
% clarify the approach of this review
With respect to XAS, in this review we focus on the low energy range
of the K edge, the pre-edge features
\cite{Westre:1997_JACS,Yamamoto:2008_XRS,DeGroot:2009_JPCM}. These
features are enhanced by collecting the fluorescence channel across
the absorption edge with a small energy bandwidth, as obtained via a
wavelength dispersive spectrometer (WDS). This technique is nowadays
referred to high energy resolution fluorescence detected (HERFD) XAS
\cite{Glatzel:2012_JESRP}. The acronym RIXS here includes resonant
x-ray emission spectroscopy (RXES), that is, the direct RIXS of
Ref.~\onlinecite{Ament:2010_RMP} or the spectator RXES of
Ref.~\onlinecite{Kotani:2012_EPJB}. In addition, the initial and
intermediate core hole states are also reported for clarity. The RIXS
done by collecting the K$\alpha_1$ emission line is denoted as
1$s$2$p_{3/2}$ RIXS. Whe refer then to x-ray emission spectroscopy
(XES) as the fluorescence yield measured after photoinization and
scanned via a WDS. The presentation of XES and RIXS follows previous
reviews
\cite{DeGroot:2001_ChemRev,Glatzel:2005_CCR,Glatzel:2009_SRN,Bergmann:2009_PR,Glatzel:2012_JESRP,Glatzel:2013_book}
by extending the applicability to magnetic semiconductors. The
specific case of non-resonant inelastic x-ray scattering
\cite{Rueff:2010_RMP}, the x-ray Raman scattering (XRS), is not
treated here. XRS permits measuring the K-edge of light elements (\eg
C, N, O) with hard x-rays \cite{Huotari:2012_JSR,Wernet:2004_S}. A
possible application of XRS is the study of doping mechanism with
shallow impurities (as the case of co-doping in magnetic
semiconductors), but it is currently not applicable to dilute
systems. The very low cross section limits the application of XRS. XES
and RIXS are also extensively employed in the soft x-ray energy range
\cite{Gelmukhanov:1999_PR,Kotani:2001_RMP,Ament:2010_RMP}. One
relevant application is the element-selective mapping of the valence
and conduction bands \cite{Preston:2008_PRB,Luning:2008_CRP}. RIXS is
also often used to study collective excitations in systems with
long-range order. The analysis of the energy dispersion as a function
of the momentum transfer permitted identifying a two-directional
modulation in the charge density of high-temperature superconductors
\cite{Ghiringhelli:2012_Sci} and the magnon dispersion
\cite{Braicovich:2010_PRL}. Another study, using soft x-rays at the L
edge of Cu in a quasi one-dimensional cuprate (Sr$_2$CuO$_3$), proved
the existence of long-sought orbitons
\cite{Schlappa:2012_Nat}. Reviewing RIXS employed to probe the
dispersion of quasiparticles and their fractionalization is beyond our
present scope. The interested reader can refer to
Ref.~\onlinecite{Ament:2010_RMP} (and references therein).
\\
% summary of the paper
The paper is organised as follows. We start by giving some elements of
the RIXS theory and present the Kramers-Heisenberg formula in
\S~\ref{sec:theory}. This is followed by an overview of current
methods employed in calculating x-ray spectra
(\S~\ref{sec:calculations}). The experiment and the required
instrumentation to perform XES and RIXS are presented in
\S~\ref{sec:exp}. The features of a RIXS intensity plane are then
discussed in \S~\ref{sec:rixs-plane}. The information content of the K
emission lines is described in \S~\ref{sec:xes}, with a focus on 3$d$
TM valence-to-core XES. The K$\beta$ core-to-core transitions as an
indirect probe of the local spin moment are presented in
\S~\ref{sec:kbeta} with a selected application to the study of
\ddms{Ga}{Mn}{Mg}{N}. This selectivity permits collecting spin- and
site-selective XAS (\S~\ref{sec:spin-xas}). A combination of RIXS with
magnetic circular dichroism (RIXS-MCD) is presented in
\S~\ref{sec:rixs-mcd}. Finally, in \S~\ref{sec:final}, our views on
future developments of the technique in the study of magnetic
semiconductors are given.\\
%
%========================================%
\section{Kramers-Heisenberg formalism}
\label{sec:theory}
\begin{figure}[!hbt]
\begin{center}
  \includegraphics[width=\MyFigSize]{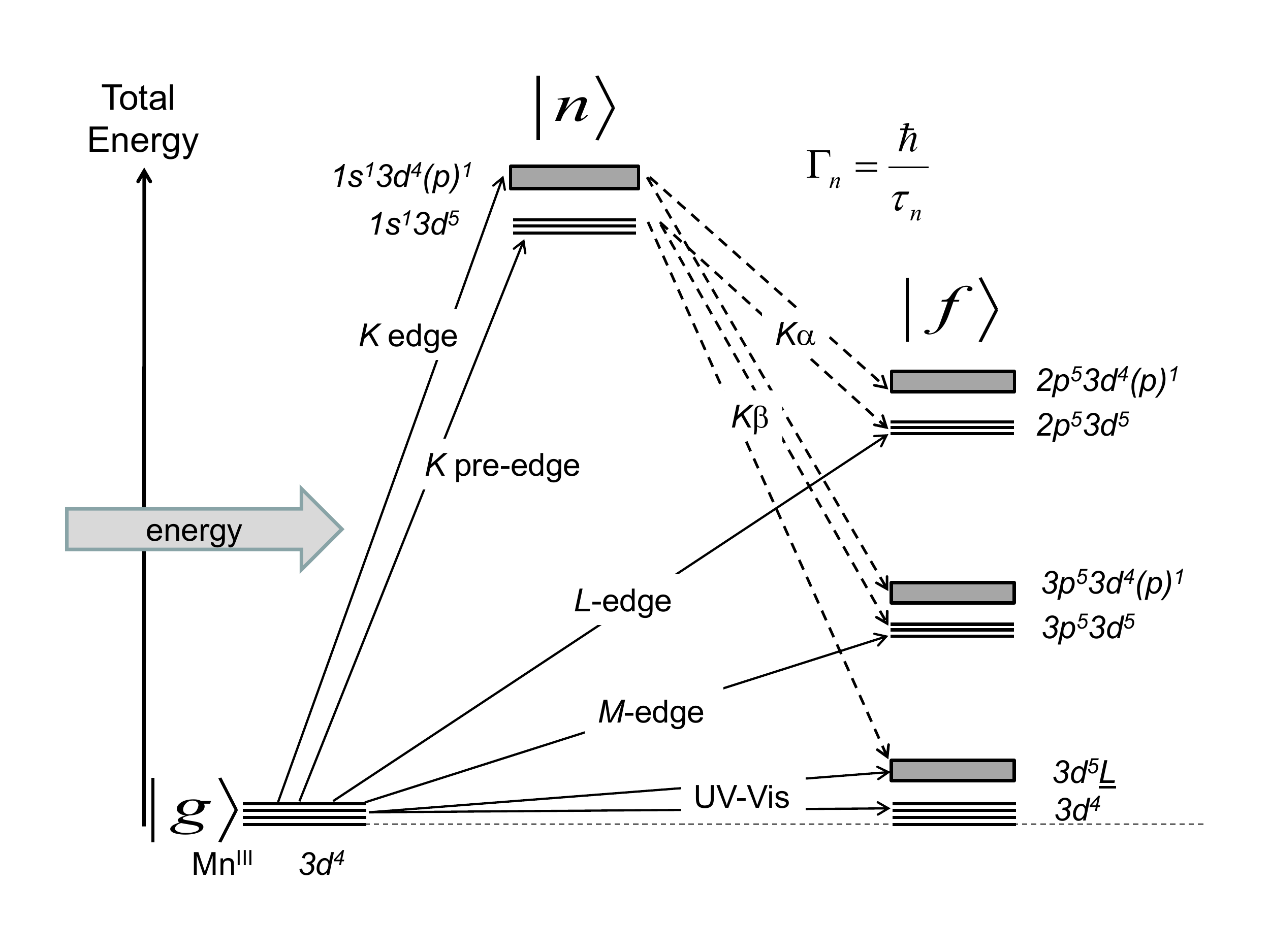}
  \caption{Total energy diagram of a system containing Mn in formal
    oxidation state III. The energy levels for the ground ($\ket{g}$),
    intermediate ($\ket{n}$) and final ($\ket{f}$) states are
    approximated using atomic configurations and a single-particle
    picture of the electronic transitions. Closed shells are omitted
    for clarity. Black lines indicate that each configuration
    corresponds to a collection of several many-body
    states. Rectangles symbolise a band state, where the notation {\em
      (p)} represents mixed states between the atomic 4$p$ level and
    the bands of the solid. The notation {\em \underline{L}}
    corresponds to a hole created on a ligand orbital. Solid-line
    arrows indicate the absorption edge that allows the corresponding
    excited state to be reached directly, \ie in one step. Dashed-line
    arrows indicate the emission line that allows excited states of
    lower energies to be reached in a second step, \ie after a core
    hole has been primarily created. Inspired from Figure 2.3 in
    Ref.~\onlinecite{Glatzel:2013_book}.}
  \label{fig:Mn_energy_dia}
\end{center}
\end{figure}

%========================================%
%
We present in the following a brief introduction to the theory of
x-ray emission spectroscopy. A more comprehensive treatment of the
theory is available in recent review papers and books: Gel'mukhanov
and \AA gren \cite{Gelmukhanov:1999_PR,Agren:2000_JERSP} (molecules),
de Groot and Kotani \cite{DeGroot:2008_book} (hard and soft x-rays in
condensed matter), Rueff and Shulka \cite{Rueff:2010_RMP} (high
pressure applications) and Ament {\em et al.} \cite{Ament:2010_RMP}
(elementary excitations in solid state physics).
\begin{figure}[!hbt]
  \begin{center}
    \includegraphics[width=\MyFigSize]{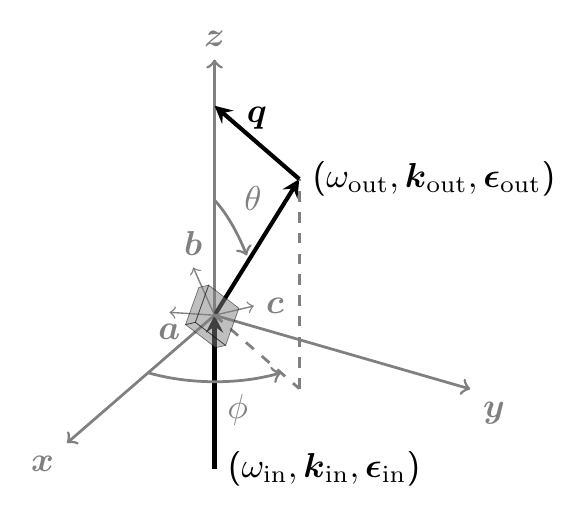}
    \caption{Scattering geometry for photon-in/photon-out
      spectroscopy. For $\phi = 0$, $\vect{\epsilon}_{\rm in}$ points
      along $\vect{x}$. The gray box represents the sample as a single
      crystal, where the cell vectors ($\vect{a}$, $\vect{b}$,
      $\vect{c}$) are not oriented with respect to the surface in
      order to emphasise the polarization effects.}
    \label{fig:scat-geo}
  \end{center}
\end{figure}
X-ray emission is a secondary process that occurs after creation of a
vacancy in an inner-shell of the element of interest. In most
applications this vacancy is created after photoexcitation and x-ray
emission becomes a photon-in/photon-out process and therefore an x-ray
scattering phenomenon. Alternatives to photoexcitation exist (\eg ion
or electron bombardement, radioactive isotopes) but the theoretical
treatment in these cases only requires minor adjustments with respect
to the following considerations. In the general description of a
scattering process (\cf figure~\ref{fig:scat-geo} for the scattering
geometry), a photon of energy $\hbar \oin$, wave vector $\vect{k}_{\rm
  in}$ and unit polarization vector $\vect{\epsilon}_{\rm in}$ is
scattered by the sample with ground state eigenfunction $\ket{g}$. A
photon is emitted into a solid angle $d \Omega$ described by a polar
angle $\theta$ and azimuthal angle $\phi$. The scattered photon has
energy $\hbar \oout$, wave vector $\vect{k}_{\rm out}$ and
polarization $\vect{\epsilon}_{\rm out}$. The energy $\hbar \omega =
\hbar(\oin- \oout)$ and momentum $\hbar \vect{q} = \hbar
(\vect{k}_{\rm in} - \vect{k}_{\rm out})$ are transferred to the
sample that consequently makes a transition from the ground state
$\ket{g}$ with energy $E_g$ to the final state $\ket{f}$ with energy
$E_f$. The derivation of the double differential scattering
cross-section (DDSCS), $d^2\sigma/(d\Omega d\hbar\oout)$, by means of
second-order perturbation treatment can be found in many textbooks
(\eg Sch\"ulke \cite{Schulke:2007_book} and Sakurai
\cite{Sakurai:1967_book}). The x-ray electromagnetic field is
represented by its vector potential $\vect{A}$. By neglecting the
interaction of the magnetic field with the electron spin, the
interaction Hamiltonian is written (in SI units) as
\cite{Ament:2010_RMP}
\begin{equation}\label{eq:hint}
H_{\rm int} = \frac{e^2}{2m} \sum_j \vect{A}(\vect{r}_j)^2 + \frac{e}{m} \sum_j \vect{p}_j \cdot \vect{A}(\vect{r}_j)
\end{equation}
where $\vect{p}_j$ is the momentum of the $j$-th target electron. The
transition probability is given by the golden rule
$\bra{s_2}T\ket{s_1}$ where $T$ is a transition operator connecting
two eigenstates. The term containing $\vect{A}^2$ does not involve the
creation of a photon-less intermediate state and can therefore be
described as a one-step scattering process (first order perturbation
theory term). It gives rise to non-resonant scattering and can, apart
from a few exceptions, not be used for an element-selective
spectroscopy. The non-resonant term accounts for elastic Thomson
scattering and inelastic Raman and Compton scattering. Inelastic
scattering may give an element-selective signature if the energy
transfer corresponds to an absorption edge
\cite{Schulke:2007_book} (the case of XRS).\\
The term containing $\vect{p} \cdot \vect{A}$ contributes to the
second order perturbation term and causes annihilation of the incoming
photon and thus the creation of an intermediate state that lives for a
time $\tau$ and decays upon emission (creation) of a photon. The
technique is element-selective if the intermediate state can be
represented by an electron configuration that contains a hole in a
core level of the element of interest. As in example, for Mn this
would be the levels 1$s$, 2$s$, \ldots, 3$p$ (\cf
figure~\ref{fig:Mn_energy_dia}).
We refer to the resonant term as the Kramers-Heisenberg cross
section. This term governs x-ray absorption (when considered as
coherent elastic forward scattering) and x-ray emission including all
resonant scattering processes. The interaction terms can be treated
separately, assuming implicitly that the experimental conditions are
chosen such that one term dominates. Other terms and interference with
them are not taken into account. Removing unimportant factors, the
essential part of the RIXS spectrum can be described in the following
form \cite{Schulke:2007_book}
\begin{eqnarray}\label{eq:kh}
\sigma(\oin, \oout) & = & r_e^2 \frac{\oout}{\oin} \sum_f \left| \sum_n \frac{\bra{f} T_{\rm out}^{*} \ket{n} \bra{n} T_{\rm in} \ket{g}}{E_g-E_n+\hbar\oin-i\Gamma_n/2} \right|^2 \nonumber \\
& \times & \delta(E_g-E_f+\hbar\oin-\hbar\oout)
\end{eqnarray}
where $r_e$ is the classical electron radius and $T$ the transition
operators ($T = \sum_j (\vect{\epsilon}\cdot\vect{p}_j)
e^{i\vect{k}\cdot\vect{r}_j}$). $\Gamma_n$ denotes the spectral
broadening due to the core hole lifetime of the intermediate state
$\ket{n}$ as a result of the Auger and radiative decays of the core
hole. The lifetime is often assumed constant for a given subshell core
hole. In order to account for the finite lifetime of the final states,
the energy-conservation $\delta$-function can be broadened into a
Lorentian of full width at half maximum $\Gamma_f$:
$\frac{\Gamma_f/2\pi}{(E_g-E_f+\hbar\oin-\hbar\oout)^2+\Gamma_f^2/4}$. A
final approximation that is employed for practical calculation of the
Kramers-Heisenberg cross section is the expansion to the second order
of the transition operators. This leads to $T \approx
\vect{\epsilon}\cdot\vect{r} + \frac{i}{2}
(\vect{\epsilon}\cdot\vect{r})(\vect{k}\cdot\vect{r})$ and corresponds
to a description of the cross section in terms of dipole (E1) and
quadrupole (E2) transitions only.\\
%
%===============================================%
\subsection{Kramers-Heisenberg equation for XES}
\label{sec:theory-xes}
%===============================================%
%
The Kramers-Heisenberg equation (Eq.~\ref{eq:kh}) is the basis for
x-ray absorption and emission spectroscopy. We note that this view is
at odds with some publications where the Kramers-Heisenberg equation
is only applied to excitations just above ($\approx $ 0-20 eV) the
Fermi level. Such excitations are referred to as resonances. However,
also x-ray emission after photoionization has to be treated using the
formalism of Eq.~\ref{eq:kh} and there is no fundamental difference
between excitations close to or well above the Fermi
level. Interference effects may be more likely to be important just
above an absorption edge but this does not suffice for a clear
distinction.\\
\begin{figure}[!hbt]
\includegraphics[width=\MyFigSize]{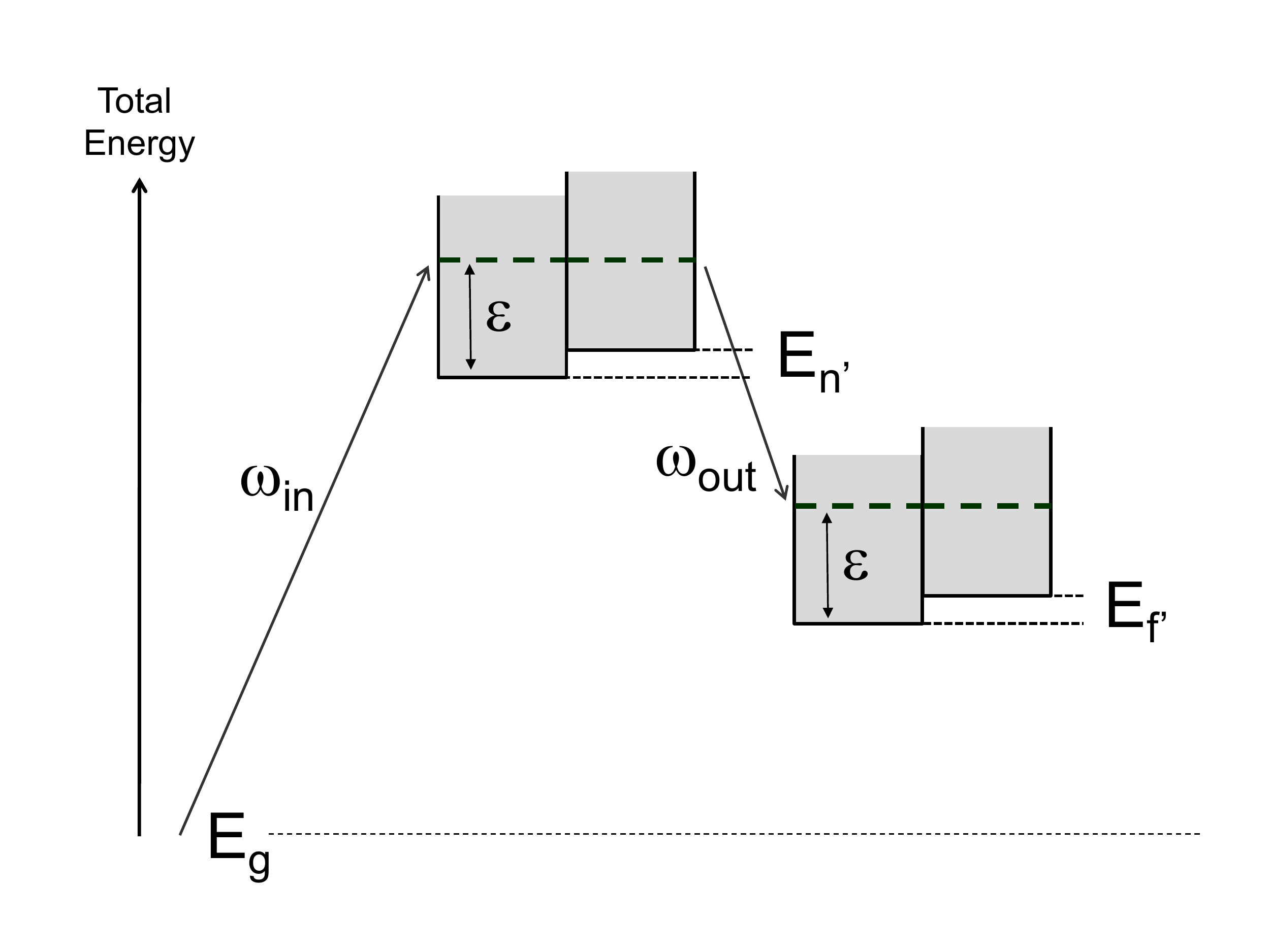}
\caption{Simplified total energy diagram for XES. A continuum of
  intermediate and final states is reached whose energies depend on
  the photoelectron kinetic energy $\varepsilon$. Several of these
  pairs of states may exist (with different $n^\prime$, $f^\prime$ and
  $\varepsilon$) that need to summed up for a full treatment of the
  x-ray emission process after photo excitation.}
\label{fig:nonresxes}
\end{figure}
The cross section changes dramatically within the first tens of eV
above an absorption edge and approaches a $\frac{1}{E^{3}}$
dependence. In this range, the photoelectron (described by its energy
$\varepsilon$) does not interact with the remaining ion; the
intermediate and final states with their energies are written as
\cite{Ljungberg:2010_PRB} $\bra{n}=\bra{n^\prime}\bra{\varepsilon}$,
$E_n=E_{n^\prime}+\varepsilon$ and
$\bra{f}=\bra{f^\prime}\bra{\varepsilon}$, $E_f =
E_{f^\prime}+\varepsilon$ (\cf figure \ref{fig:nonresxes}). The
photoelectron does not change its energy upon the radiative decay of
the ion. One thus obtains, for each ionic intermediate and final
state, $n^\prime$ and $f^\prime$, an infinite number of states
characterised by the kinetic energy of the
photoelectron. Eq.~\ref{eq:kh} then becomes
\begin{eqnarray}\label{eq:kh_simple1}
\sigma(\oin, \oout) \propto \nonumber\\
\sum_{f^\prime} \sum_{n^\prime} \int_\varepsilon \left| \frac{ \bra{f^\prime}\bra{\varepsilon} T_{\rm out}^{*} \ket{\varepsilon}\ket{n^\prime} \bra{n^\prime}\bra{\varepsilon} T_{\rm in} \ket{g}}{E_g - (E_{n^\prime} + \varepsilon) + \hbar\oin - i\Gamma_n/2} \right|^2 \nonumber\\
\times \frac{\Gamma_f/2\pi}{(E_g - (E_{f^\prime} + \varepsilon) + \hbar\oin - \hbar\oout)^2 + \Gamma_f^2/4}
\end{eqnarray}
We assume constant absorption and emission matrix elements for each
$n^\prime$, $f^\prime$, \ie independent of the photoelectron kinetic
energy $\varepsilon$ (which is justified over a small energy range),
and we obtain
\begin{eqnarray}\label{eq:kh_simple2}
\sigma(\oin, \oout) \propto \nonumber\\
\sum_{f^\prime} \sum_{n^\prime} \left| \bra{f^\prime} \bra{\varepsilon} T_{\rm out}^{*} \ket{\varepsilon} \ket{n^\prime} \bra{n^\prime} \bra{\varepsilon} T_{\rm in} \ket{g} \right|^2 \nonumber\\
\times \int_\varepsilon \frac{1}{(E_g - (E_{n^\prime} + \varepsilon) + \hbar\oin)^2 + \Gamma_n^2/4} \nonumber\\
\times \frac{\Gamma_f/2\pi}{(E_g - (E_{f^\prime} + \varepsilon) + \hbar\oin-\hbar\oout)^2 + \Gamma_f^2/4}
\end{eqnarray}
The integral over $\varepsilon$ is a convolution of two Lorentzian
functions which gives a Lorentzian as a function of $\oout$ with width
$\Gamma_n+\Gamma_f$ which is the known result for non-resonant
fluorescence spectroscopy. A broad energy bandwidth for the incident
beam will result in a larger range of photoelectron kinetic energy but
not influence the width of the convoluted Lorentzian. Hence, the
spectral broadening is independent of the incident energy
bandwidth. This opens the door to experiments using non-monochromatic
radiation with a bandwidth of $\Delta E < 100$~eV (pink beam) at
synchrotron radiation sources or free electron
lasers \cite{Kern:2013_S}.\\
The spectral shape does not depend on $\varepsilon$ as long as the
same set of intermediate states $n^\prime$ is reached. This may change
if the incident energy suffices to create more than one core hole (\cf
Ref.~\onlinecite{Hoszowska:2013_JESRP} and references therein). One
example is the KL-edge where one incident photon creates a hole in the
K- and L-shell. This may significantly alter the x-ray emission
spectral shape. It is therefore important to choose the incident energy
below the edge of multi-electron excitations, if possible.
%
%===============================================%
\section{Approaches to the calculations}
\label{sec:calculations}
%===============================================%
%
In this section we will present a short overview of the methods
currently employed to calculate the experimental spectra. The
theoretical simulation is an important tool for the experimentalist
who needs to analyse the collected spectra and to plan new
experiments. We can roughly separate the various approaches to the
calculations of inner-shell spectra into two main philosophies, that
we can refer to as: many-body atomic picture and single-particle
extended picture. They are based, respectively, on ligand field
multiplet theory (LFMT) and density functional theory (DFT).\\
In LFMT one first considers a single ion and writes its wavefunction
as a single or linear combination of Slater determinants of atomic
one-electron wavefunctions. The chemical environment is then
considered by empirically introducing the crystal field splittings and
the orbital mixing. A detailed description of LFMT can be found in
textbooks \cite{Bransden:1983_book,Figgis:2000_book,DeGroot:2008_book}
or topical reviews
\cite{Griffith:1957_QRCS,DeGroot:2001_ChemRev,DeGroot:2005_CCR}, while
a tutorial-oriented description of the calculations was given by van
der Laan \cite{VanDerLaan:2006_book}. The codes currently in use are
those developed by Cowan \cite{Cowan:1968_JOSA,Cowan:1981_book} in the
sixties and extended by Thole in the eighties (\cf
Ref.~\onlinecite{VanderLaan:1997_JESRP} for a technical
overview). Recently, a user friendly interface, {\sc ctm4xas}
\cite{Stavitski:2010_M}, has permitted a larger community accessing
such calculations. The advantage of this approach is that the core
hole is explicitely taken into account and multi-electron effects are
calculated naturally by applying multiplet theory. The obvious problem
with this approach is that the chemical environment is only considered
empirically.\\
In the DFT-based approach, a simplified version of the Schr\"odinger
equation is solved either for a cluster of atoms centered around the
absorbing one (real space method) or using periodic boundary
conditions (reciprocal space method). This means that the electronic
structure is calculated {\em ab initio}, without the need of empirical
parameters, and the results depend on the level of approximation
employed. Among the large number of presently used codes, the most
common techniques are: multiple scattering theory (\eg {\sc feff9}
\cite{Rehr:2010_PCCP,Rehr:2009_CRP}, {\sc fdmnes}
\cite{Joly:2001_PRB,Bunau:2009_JPCCM} and {\sc mxan}
\cite{Benfatto:2001_JSR}), full potential linearised augmented plane
wave, FLAPW (\eg {\sc Wien2k}
\cite{Schwarz:2003_CMS,Pardini:2011_CPC}), projector augmented-wave
method, PAW (\eg {\sc quantum-espresso}
\cite{Giannozzi:2009_JPCM,Gougoussis:2009_PRB,Bunau:2013_PRB}, {\sc
  gpaw} \cite{Enkovaara:2010_JPCM,Ljungberg:2011_JESRP}, {\sc bigdft}
\cite{Genovese:2008_JCP}) and time-dependent DFT (\eg {\sc orca}
\cite{Neese:2012_RCMS,DeBeer:2010_IC}). The advantage in the DFT
approach is that the theoretical framework is well established and
numerous groups work on evaluating and improving the level of theory,
\ie the exchange-correlation functionals or the basis sets. However,
DFT is a theory to calculate the ground state electronic structure
which is {\em a priori} incompatible with inner-shell
spectroscopy. Furthermore, in its basic implementation, DFT calculates
one-electron transitions which are insufficient when the inner-shell
vacancy gives a pronounced perturbation of the electronic structure,
resulting in important many-body effects. These shortcomings have been
addressed within DFT and considerable progress has been made
\cite{Onida:2002_RMP}.\\
The decision on which approach is most suitable for the problem at
hand can be based on the degree of localization of the orbitals that
are assumed to be involved in the transitions. The K absorption main
edge in 3$d$ transition metals is often modeled using DFT. The
pre-edge requires a mixture of atomic and extended view and therefore
only in a few favorable cases a good understanding of the pre-edge
features has been achieved. The L-edges of rare earths and 5$d$
transition metals require an extended approach. However, 2$p$ to 4$f$
transitions that form the L pre-edge in rare earths are highly
localised and an atomic approach is very successful. The K$\beta$ main
line emission in 3$d$ transition metals involve atomic
orbitals. Multiplet theory can therefore reproduce the spectral shape
to high accuracy. In contrast, the valence-to-core lines involve
molecular orbitals that are mainly localised on the ligands and a
one-electron DFT approach is therefore very succesful in reproducing
the spectra.\\
It is often illuminating to apply a very simplified approach to
simulate an experimental result, as it permits assessing what
interactions and effects are relevant. As an example, if one neglects
interference effects, the core hole potential and multi-electron
transitions, it is possible to drastically simplify the
Kramers-Heisenberg formula (Eq.~\ref{eq:kh}) for the case of
valence-to-core RIXS and obtain an expression in terms of the angular
momentum projected density of states \cite{Jimenez-Mier:1999_PRB}
\begin{equation}\label{eq:dos2rixs}
\sigma(\oin, \oout) \propto \int_\varepsilon \frac{\rho(\varepsilon)\rho^\prime(\varepsilon + \oin - \oout)}{(\varepsilon - \oout)^2 + \Gamma_n^2/4} d\varepsilon
\end{equation}
where $\rho$ and $\rho^\prime$ are, respectively, the occupied and
unoccupied density of states, $\Gamma_n$ the lifetime broadening of
the intermediate state. This approach has been demonstrated valid in
describing the \vtc{RIXS} spectra of 5$d$ transition metal systems
\cite{Smolentsev:2011_PRB,Garino:2012_PCCP}. A similar approach but
partly considering the core hole potential and the radial matrix
element was recently implemented in the {\sc feff9} code
\cite{Kas:2011_PRB}.\\
The combination of an extended picture with full multiplet
calculations is the holy grail in theoretical inner-shell
spectroscopy. The progress in recent years has been impressive to the
great benefit of the experimentalists who are gradually getting a
better handle on analyzing their data
\cite{Mirone:2000_PRB,Uldry:2012_PRB,Mirone:2012_CPC}. A promising
method is to make use of maximally localised Wannier functions
\cite{Marzari:2012_RMP} as directly obtained from DFT calculations. If
one extracts the Wannier orbitals in the bands near the Fermi level,
is then possible to calculate the spectra via LFMT
\cite{Haverkort:2012_PRB}. However, this method is still an
approximate solution of the problem. A more rigorous treatement was
proposed in the framework of the multi-channel multiple scattering
(MCMS) theory \cite{Natoli:1990_PRB} (recently revised in
Ref.~\onlinecite{Natoli:2012_JPCM}). The MCMS method has been
successfully applied in simulating the L$_{2,3}$ XAS spectra of Ca
\cite{Kruger:2004_PRB} and Ti \cite{Kruger:2010_PRB} and could be
easily extended to XES and RIXS.\\
%
%======================%
\section{Experimental set-up}
\label{sec:exp}
%======================%
%
\begin{figure*}[!hbt]
\includegraphics[width=\MyFigSizeStar]{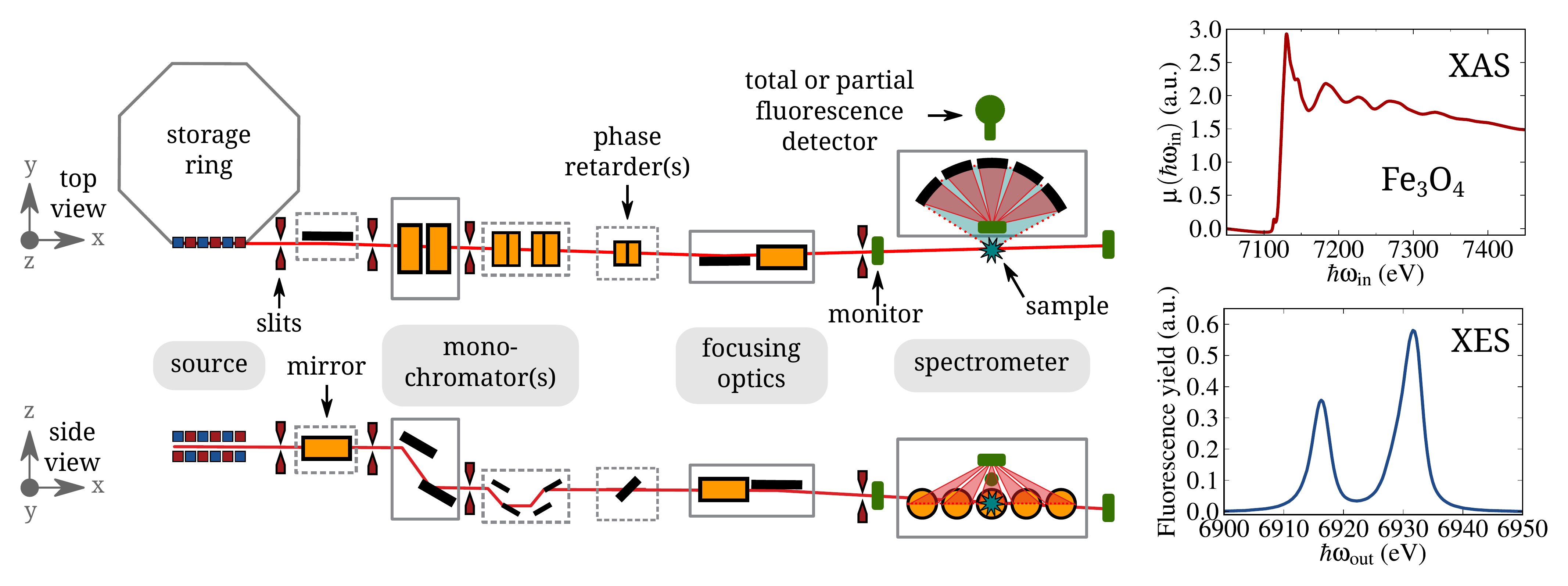}
\caption{Schematic view (top/side) of a generic hard x-ray
  RIXS-dedicated beamline: optics and experimental stations (\cf
  description in the text). The represented objects are not in scale
  and do not represent a given technical design. Right panels show an
  example of XAS (scanning the incoming energy, $\hbar \oin$, top) and
  XES (scanning the outgoing energy, $\hbar \oout$, bottom) spectra of
  Fe$_3$O$_4$ powder obtained at the Fe K-edge.}
\label{fig:rixs-exp}
\end{figure*}
Before presenting a selection of applications of the technique, we
describe how a combined XAS/XES experiment is performed on a generic
synchrotron radiation beamline. This is schematically illustrated in
figure~\ref{fig:rixs-exp}. The synchrotron radiation is produced in the
storage ring via an undulator, bending magnet or wiggler (source). A
first collimating mirror, run in total reflection geometry, is usually
used to reduce the heat load, collimate the beam and remove the higher
harmonics. The beam is then monochromatised by a double single crystal
monochromator (cryogenically cooled); typically two pairs of crystals
are employed: Si(111) or Si(311), giving an intrinsic (without taking
into account the beam divergence) resolving power, \rp, of 7092 and
34483 \cite{Matsushita:1983_book}, respectively. The monochromatic
beam is then focused to the sample via a focusing system, typically,
two bent mirrors in Kirkpatrick-Baez geometry
\cite{Kirkpatrick:1948_JOSA}, that is, working in glancing incidence
(around 3 mrad), one focusing horizontally and the second one,
perpendicular to the previous, focusing vertically. A given number of
slits (vertical and horizontal) is also inserted in the beam path to
clean for aberrations and reduce the divergence. In addition, the
beamline optics can be complemented with a second monochromator or
phase retarders. The second monochromator, typically a channel cut in
four crystals configuration \cite{DuMond:1937_PR}, is used to improve
the energy resolution. The phase retarders \cite{Giles:1995_RSI},
typically thin diamond crystals put in diffraction conditions, permit
tuning the polarization of the x-ray beam. In fact, apart for helical
undulators, the x-ray beam is linearly polarised in the orbital plane
and the phase retarders are required for generating circularly
polarised light (left and right) or linearly polarised in the
vertical plane.\\
In the experimental station, the equipment is built around the sample
stage (figure~\ref{fig:rixs-exp}). The main elements consist in x-ray
detectors for monitoring the incoming and transmitted beam and
measuring the fluorescence emitted by the excited sample. For hard
x-rays, the sample environment does not require a vacuum chamber and
it is quite versatile: a goniometer permits aligning the sample in
three dimensions plus hosting additional equipments (\eg cryostat,
furnace, magnet or chemical reactor). For bulk samples, the XAS
(absorption coefficient, $\mu(\hbar\oin)$) is measured directly via
the intensity of the incoming ($I_{\rm in}$) and transmitted beam
($I_{\rm out}$), according to the Beer-Lambert law: $I_{\rm
  out}=I_{\rm in}e^{-\mu(\hbar\oin)x}$, where $x$ is the sample's
thickness. For dilute species, $\mu$ cannot be measured directly and a
secondary process (yield) has to be employed, assuming that the
absorption cross section is proportional to the number of core holes
created. The secondary processes can be either the collection of the
electron yield \cite{Erbil:1988_PRB} or the fluorescence yield
\cite{Jaklevic:1977_SSC}. We will not treat the electron yield here,
but focus on the fluorescence yield (FY) because this gives access to
a photon-in/photon-out spectroscopy, as XES and RIXS. Usually, the
FY-XAS is collected either without energy resolution (total FY) or
with an energy dispersive solid state detector (SSD) as an array of
high purity germanium elements or silicon drift diodes. For linearly
polarised synchrotron radiation (with $\vect{\epsilon}_{\rm in}$ along
$\vect{x}$, \cf figure~\ref{fig:scat-geo}, as in standard experiments)
the Thomson (elastic), Compton and Raman (inelastic) scattering have
an angular dependence of $sin^2(\phi)+cos^2(\theta)cos^2(\phi)$ (\cf
figure~\ref{fig:scat-geo}) while the fluorescence emitted by the sample
is isotropic (in a standard geometry and not considering polarization
effects, \cf Ref.~\onlinecite{Bianchini:2012_JSR} for the full
expression), thus the fluorescence detectors are usually put at 90
degrees on the polarization plane to minimise the background due to
scattering (\cf figure~\ref{fig:rixs-exp}). SSD detectors permit a
typical energy resolution of 150-300 eV (\rp~$\approx$~50). This low
energy resolution combined with a low saturation threshold is a
drawback for measuring dilute species in strong absorbing matrices as
DMS/CMS. In fact, the weak signal of interest is very often sitting on
the strong background coming from the low-energy tail of the Thompson
and Compton scattering or overlapping with the fluorescence lines of
the other elements contained in the matrix. For thin films deposited
on a substrate, a workaround for collecting a clean fluorescence
signal is to work in a combined grazing incidence and grazing exit
geometry \cite{Maurizio:2009_RSI} but this has the drawback of fixing
the experimental geometry and it is not suitable for single crystals
where it is important also to work with the polarization axis laying
out of the sample surface. An increased energy resolution
(\rp~$\approx$~1000) can be obtained with charged coupled devices
\cite{Fourment:2009_RSI} or microcalorimetric arrays
\cite{Uhlig:2013_PRL} used in energy resolving mode. However, the
complexity of these detectors (especially in the events reconstruction
algorithms) and the very quick saturation for calorimeters, limits
their application on standard spectroscopy beamlines.\\
In order to overcome these limitations and to collect XES, RIXS and
HERFD-XAS, a wavelength dispersive spectrometer has to be employed
(\rp~$>$~5000). For hard x-rays, this means that Bragg's diffraction
over an analyser crystal is employed to monochromatise the emitted
fluorescence from the sample (Rowland's circle geometry). Among all
the possible diffraction geometries \cite{Pestehe:2011_JOSAA}, two
main configurations are currently in use at synchrotron facilities:
the point-to-point Johann \cite{Johann:1931_ZP} and the dispersive Von
Hamos \cite{VonHamos:1933_AP}. For both, the basic principle is that
the source (sample), the diffractor (analyser crystal) and the image
(detector) are on the Rowland circle. The first class uses spherically
bent crystals \cite{Verbeni:2005_JPCS} in combination with
one-dimension detector; the energy selection is performed by scanning
the crystal Bragg's angle and the detector over the Rowland circle. In
the second class, a cylindrically bent crystal is combined with a
position-sensitive detector; the energy dispersion is obtained without
moving the crystal and by collecting the different areas of the
detector. Without going into the details of the advantages and
disadvantages of each configuration, good performances are obtained
with an increased number of spherically bent crystals (to overcome the
small solid angle collected, $\approx$ 0.03 sr per crystal) working at
Bragg' angles close to 90 deg. As few examples of currently available
instruments, there are those dedicated to XRS
\cite{Verbeni:2009_JSR,Sokaras:2012_RSI}, medium-resolution RIXS
\cite{Glatzel:2009_SRN,Kleymenov:2011_RSI,Llorens:2012_RSI,Sokaras:2013_RSI}
and single-shot XES \cite{Szlachetko:2012_RSI,Alonso-Mori:2012_RSI}.\\
%
%==============================================================%
\section{The RIXS plane and sharpening effects}
\label{sec:rixs-plane}
\begin{figure}[!hbt]    
  \includegraphics[width=\MyFigSize]{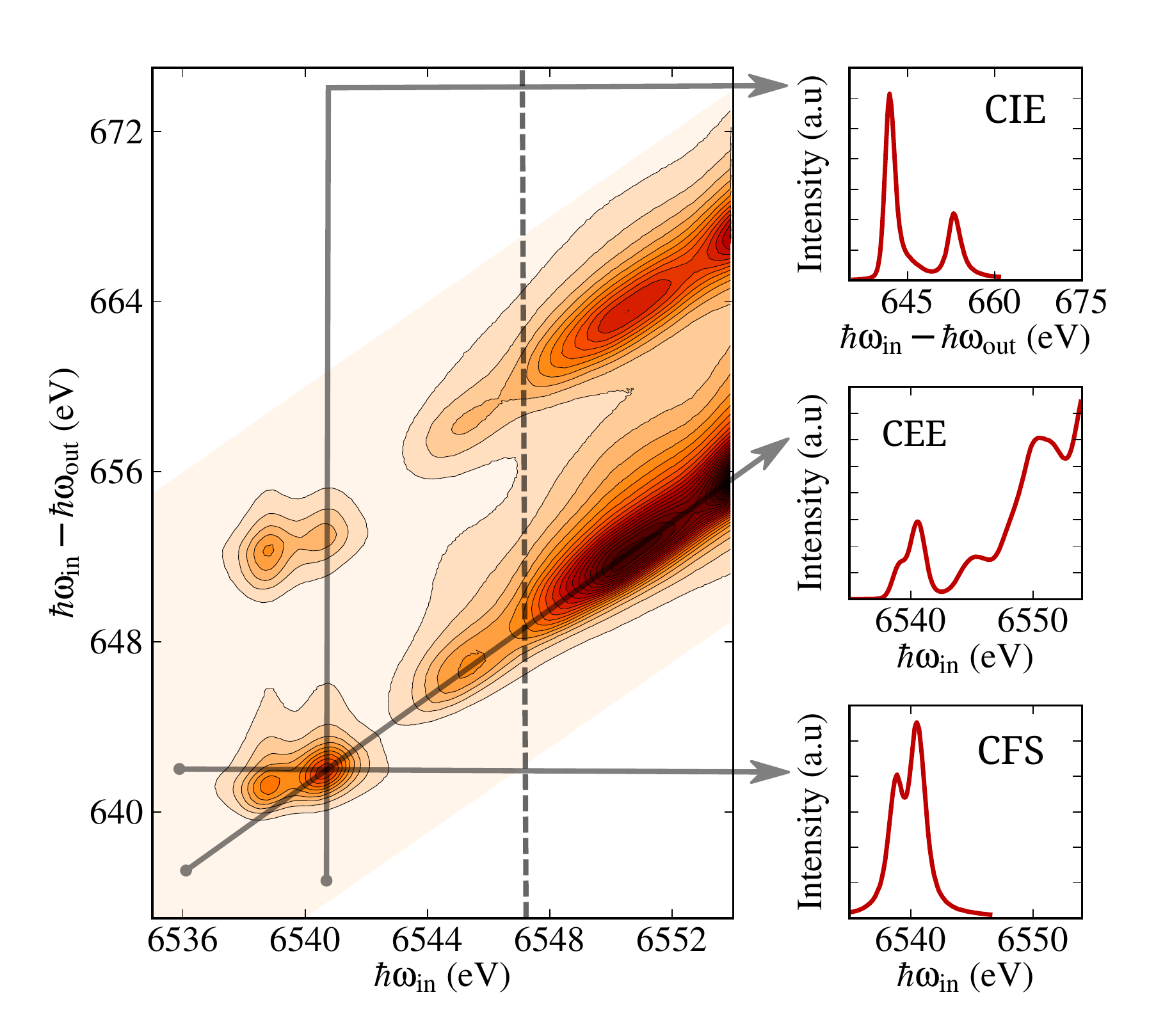}
  \caption{1$s$2$p$ RIXS intensity plane for a Ga$_{0.97}$Mn$_{0.03}$N
    thin film (\cf Ref.~\onlinecite{Stefanowicz:2010_PRB} for the
    sample's growth and characterization). The insets show line cuts
    in three main directions: at constant incident energy (CIE =
    energy loss), constant emitted energy (CEE = HERFD-XAS) and
    constant final state energy (CFS). The vertical dashed line
    indicates the absorption edge position.}
  \label{fig:rixs-plane}
\end{figure}

%==============================================================%
%
An experimental 1$s$2$p$ RIXS intensity plane is shown in
figure~\ref{fig:rixs-plane} in the incident ($\hbar\oin$) versus
transfer ($\hbar\oin-\hbar\oout$) energy axis. The transitions to
continuum (main absorption edge) appear as dispersive features along
the diagonal, while the transitions to localised states (pre-edge
features) appear as resonances at well defined positions in the
plane. The two groups of diagonal features visible in
figure~\ref{fig:rixs-plane} are vertically split by the 2$p$ spin-orbit
interaction in the final state, corresponding to the K$\alpha_1$ and
K$\alpha_2$ emission lines. A diagonal cut (constant emitted energy)
will then give the HERFD-XAS spectrum, while an integration over the
vertical direction results in a standard XANES spectrum. Considering
only the pre-edge region, a vertical cut (constant incident energy)
gives a spectrum sentitive to the spin-orbit interaction in the final
state and the exchange interaction between the intermediate and final
states. This is similar to L$_{2,3}$ edges XAS. A cut in the
horizontal direction (constant final state energy) is affected by the
spin-orbit and exchange interaction in the intermediate state only. On
the other hand, analyzing RIXS data as line scans can lead to false
interpretation of the spectral features. For example, the two pre-edge
peaks in figure~\ref{fig:rixs-plane} have an incoming energy separation
of 1.8 eV that would be underestimated (1.4 eV) if a peak-fitting
procedure is employed on the HERFD-XAS scan. This is due to the fact
the the first resonance does not lie on the diagonal of the RIXS plane
(\cf figure~\ref{fig:rixs-plane}).\\
One appreciated feature of RIXS/HERFD-XAS is a dramatic improvement in
resolving the spectral features (sharpening effect). The effect is
striking at the L$_{2,3}$ edges of 5$d$ elements when compared to
standard XANES while at K pre-edge of 3$d$ elements permits catching
fine details due to the strong reduction of the background signal. For
example, HERFD-XAS has permitted precisely following catalytic
reactions \cite{Hartfelder:2012_CST} or to reveal angular-dependent
core hole effects \cite{Juhin:2010_PRB}. The origin of the sharpening
effect was attributed to interference causing the elimination of the
core hole broadening \cite{Hamalainen:1991_PRL}. Actually, the
interference does not play a role here and the lifetime broadenings
are still present, as shown by the elongated features in the
horizontal and vertical direction of the RIXS plane (\cf
figure~\ref{fig:rixs-plane}). Without going into the details of the
difference between HERFD-XAS and standard XAS spectra, as previously
discussed by Carra {\em et al.}~\cite{Carra:1995_PRL}, it was
demonstrated that the improved resolution of the experimental spectra
can be reproduced by an apparent broadening \cite{DeGroot:2002_PRB}
\begin{equation}\label{eq:herfd-res}
\Gamma_{\rm exp} \approx \frac{1}{\sqrt{(1/\Gamma_{n})^2 + (1/\Gamma_{f})^2}}
\end{equation}
where the intermediate ($\Gamma_{n}$) and final ($\Gamma_{f}$) core
hole lifetime broadenings are taken into account.
%
%==============================================================%
\section{Valence states sensitivity of K fluorescence lines}
\label{sec:xes}
%==============================================================%
%
The macroscopic properties of semiconductors (\eg transport,
magnetism) are driven by impurities (defects) located at valence
states. Accessing the information of such states via a bulk probe,
permits then having a detailed description of the material under
study. XES can probe valence electrons either indirectly or directly,
by selecting the yield for different transitions. If one collects
core-to-core ({\em ctc}) transitions, the valence electrons are probed
indirectly, while directly for valence-to-core ({\em vtc}). The
selectivity to the electronic structure of the valence shell in
\ctc{XES} originates from screening effects (the core levels energy is
affected by the modified nuclear potential) and multiplet structure
(the spin and orbital angular momentum of the core hole strongly
couple to the valence electrons). The screening dominates for light
elements as, for example, the K$\alpha$ XES of S
\cite{AlonsoMori:2009_AC}, while the multiplet structure dominates in
the case of the K fluorescence lines of 3$d$ TMs.\\
\begin{figure}[!hbt]
\includegraphics[width=\MyFigSize]{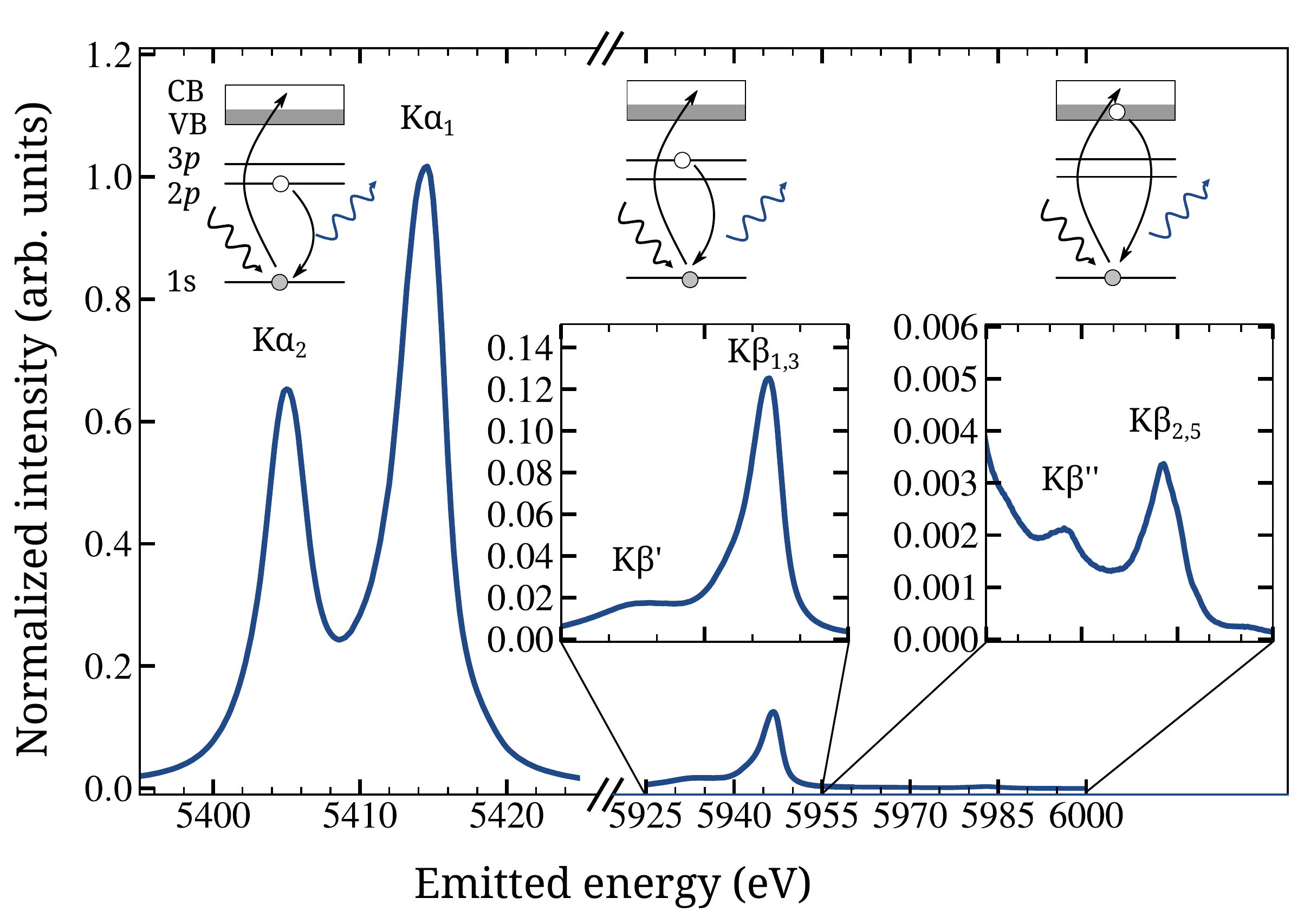}
\caption{XES spectrum of Cr$_2$O$_3$ showing the Cr K fluorescence
  lines (blue): \ctc{K$\alpha$}, \ctc{K$\beta$} and \vtc{K$\beta$}. In
  the plots the intensity is normalised to the K$\alpha_1$ maximum;
  the insets show an expanded view for \ctc{K$\beta$} and
  \vtc{K$\beta$}. The top schemes show the origin of the transitions
  after photoionization in a simplified one-electron picture. Black
  lines indicate core states (1$s$, 2$p$, 3$p$), while the rectagles
  symbolise the valence and conduction bands (VB and CB).}
\label{fig:xes-regions}
\end{figure}
A XES spectrum is given in figure~\ref{fig:xes-regions} for Cr in
Cr$_2$O$_3$. The most intense lines are the K$\alpha_1$ (K-L$_3$,
2$p_{3/2}$ $\to$ 1$s$) and K$\alpha_2$ (K-L$_2$, 2$p_{1/2}$ $\to$
1$s$), where the 2$p$ level is splitted by the strong spin-orbit
interaction. With $\approx$ 10-times smaller intensity are visible the
\ctc{K$\beta$} lines (K-M$_{2,3}$, 3$p$ $\to$ 1$s$), called
\cite{Tsutsumi:1959_JPSJ} K$\beta_{1,3}$ (main peak) and
K$\beta^\prime$ (broad shoulder at lower emitted energy). Finally, the
\vtc{K$\beta$} lines appear on the tail of the main lines with roughly
200-times smaller intensity, those are called K$\beta_{2,5}$ and
K$\beta^{\prime\prime}$. The origin and information content of
\ctc{K$\beta$} lines is given later in \S~\ref{sec:kbeta}, while here
we focus first on \vtc{K$\beta$} lines.\\
The \vtc{K$\beta$} arise from transitions from occupied orbitals a few
eV below the Fermi level (the valence band), that is, from orbital
mixed metal-ligand states of metal $p$-character to 1$s$. For this
reason, \vtc{K$\beta$} is strongly sensitive to ligand species and has
been employed in chemistry to distinguish between ligands of light
elements
\cite{Bergmann:1999_CPL,Safonov:2006_JPCB,Eeckhout:2009_JAAS,Lancaster:2011_S}
(\eg C, N, O, S). In addition, by making use of the XES polarization
dependence \cite{Glatzel:2005_CCR} is possible to study the
orientation of the lingands. For example, Bergmann {\em et al.}
\cite{Bergmann:2002_JCP} studied a Mn nitrido coordination complex in
$C_{4v}$ symmetry with five CN and one N ligand at a very short
distance (1.5~\AA). The signal arising from the nitrido molecular
orbitals was almost completely suppressed by orienting the Mn-nitrido
bond in the direction of $\vect{k}_{\rm out}$, \ie towards the crystal
analyser. Another advantage of \vtc{K$\beta$} is the possibility to
easily calculate the transitions with a molecular-orbital approach:
from early atomic \cite{Best:1966_JCP} to recent DFT methods
\cite{Pollock:2011_JACS,Gallo:2011_PCCP,Vila:2011_JPCA}. These works
demonstate that the K$\beta^{\prime\prime}$ and K$\beta_{2,5}$ are
mainly sensitive, respectively, to the ligand $s$ and $p$ states. The
K$\beta_{2,5}$ has also a strong dependece on the metal's local
symmetry \cite{Smolentsev:2009_JACS,Gallo:2013_CPC} (\eg T$_D$ \vs
O$_h$). For a more rigorous treatement, the interested reader can
refer to a recent topical review \cite{Gallo:2014_AdvMats}.\\
\begin{figure}[!hbt]
  \includegraphics[width=\MyFigSize]{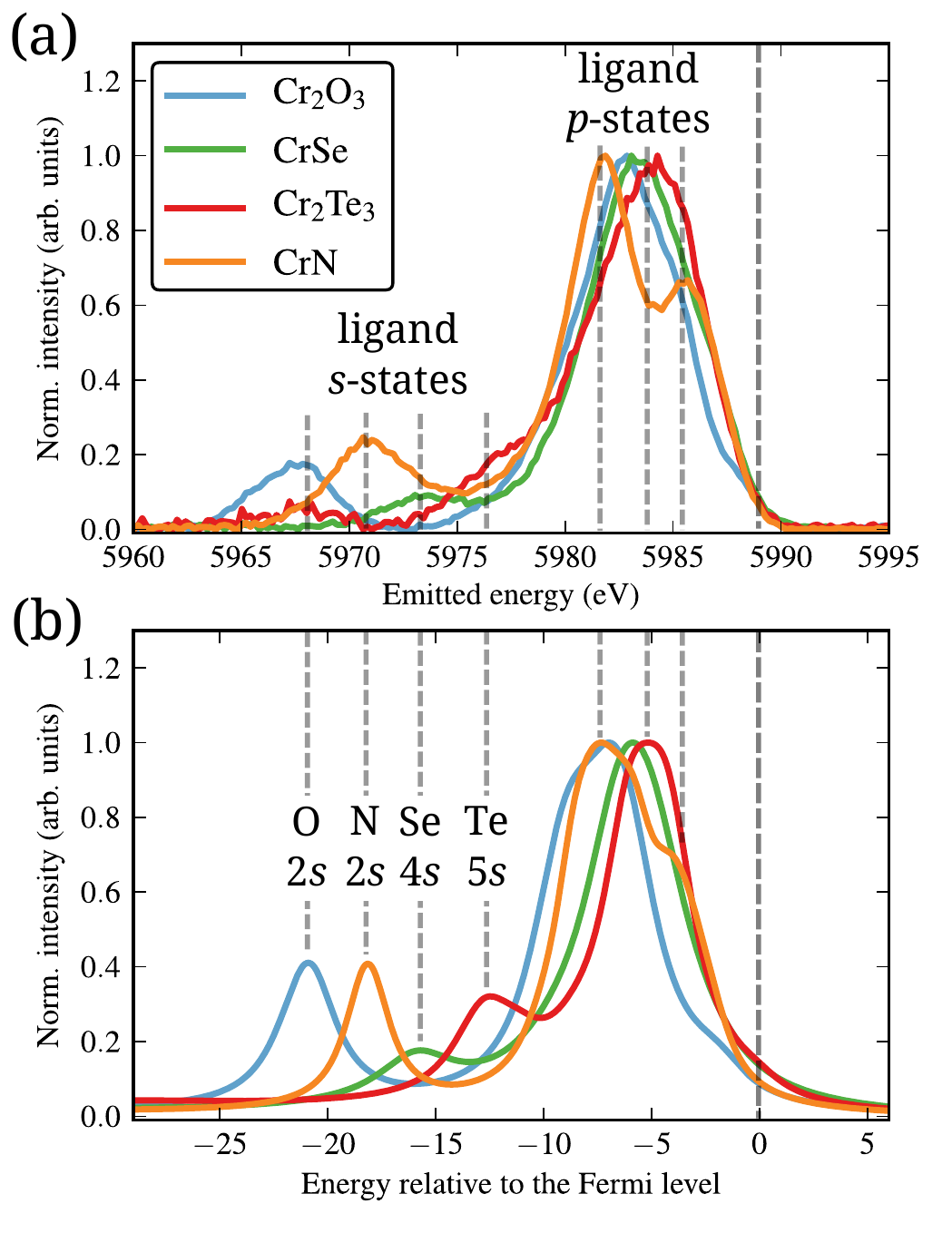}
  \caption{\vtc{K$\beta$} spectra for a selected set of Cr-based
    compounds. Top panel (a): experimental data - after removal of the
    \ctc{K$\beta$} background and normalization to the K$\beta_{2,5}$
    peak maximum - from commercially available powders. Bottom panel
    (b): {\em ab initio} simulations with the {\sc fdmnes} code (\cf
    details in the text). The ligands' states sensitivity of
    K$\beta^{\prime\prime}$ and K$\beta_{2,5}$ is outlined. Vertical
    dashed lines are a guide for the eye to compare the main spectral
    features between the simulation and the experiment.}
  \label{fig:cr-vtc-xes}
\end{figure}
In DMS, one application of \vtc{K$\beta$} is in the study of the
interaction of shallow dopants with the metal site. For example, in
\dms{Zn}{Cr}{Te} co-doped with N \cite{Kuroda:2007_NM}, where N is an
acceptor (brings holes in the valence band), by measuring the Cr
\vtc{K$\beta$} it is possible to detect when N interacts with Cr via
the clear signature of N 2$s$ levels in the Cr
K$\beta^{\prime\prime}$. This is illustrated in
figure~\ref{fig:cr-vtc-xes} where we show a selection of
\vtc{K$\beta$} spectra for commercially available Cr-based powder
compounds compared with the {\em ab initio} simulations ({\sc fdmnes}
code \cite{Bunau:2009_JPCCM}). Standard crystal structures (retrieved
from the ``Inorganic Crystal Structure Database'', FIZ Karlsruhe) are
used as input in the calculations, conducted in real space with a
muffin-tin approximation and the Hedin-Lundqvist exchange-correlation
potential. To compare with the experiment, the calculated Fermi levels
are arbitrarily shifted and the spectra are convoluted with a constant
Lorentian broadening of 2.68 eV. The origin of the features is then
attributed by selecting the projected density of states on the ligands
that overlaps with the metal $p$ one (not shown). This confirms
previous works, that is, the K$\beta^{\prime\prime}$ mainly comes from
the ligand $s$ states, while the K$\beta_{2,5}$ is from the ligand $p$
states. As shown in figure~\ref{fig:cr-vtc-xes}, the energy position
of the K$\beta^{\prime\prime}$ is very sensitive to the type of ligand
and permits identifying if a compound has an additional phase. For
example, the experimental Cr$_2$Te$_3$ spectrum shows a second
K$\beta^{\prime\prime}$ at 5967 eV, corresponding to oxygen, that is
not reproduced in the calculation. The origin of this extra feature is
then easily understood by the fact that Cr$_2$Te$_3$ is an
air-sensitive compound and, due to the measurements carried in air, it
was contaminated by oxygen. The analysis of the K$\beta_{2,5}$ is more
demanding, because its spectral features are also affected by the
local symmetry. This effect is also shown in
figure~\ref{fig:cr-vtc-xes}, where the simulated and experimental
K$\beta_{2,5}$ do not fully align. One reason resides in the fact that
the simulation takes into account only one crystallographic structure,
while the commercial powders may contain more than one crystal phase
of the same compound.
%
%===================================================================%
\section{K$\beta$ spin sensitivity via intra-atomic exchange interaction}
\label{sec:kbeta}
\begin{figure}[!hbt]
\includegraphics[width=\MyFigSize]{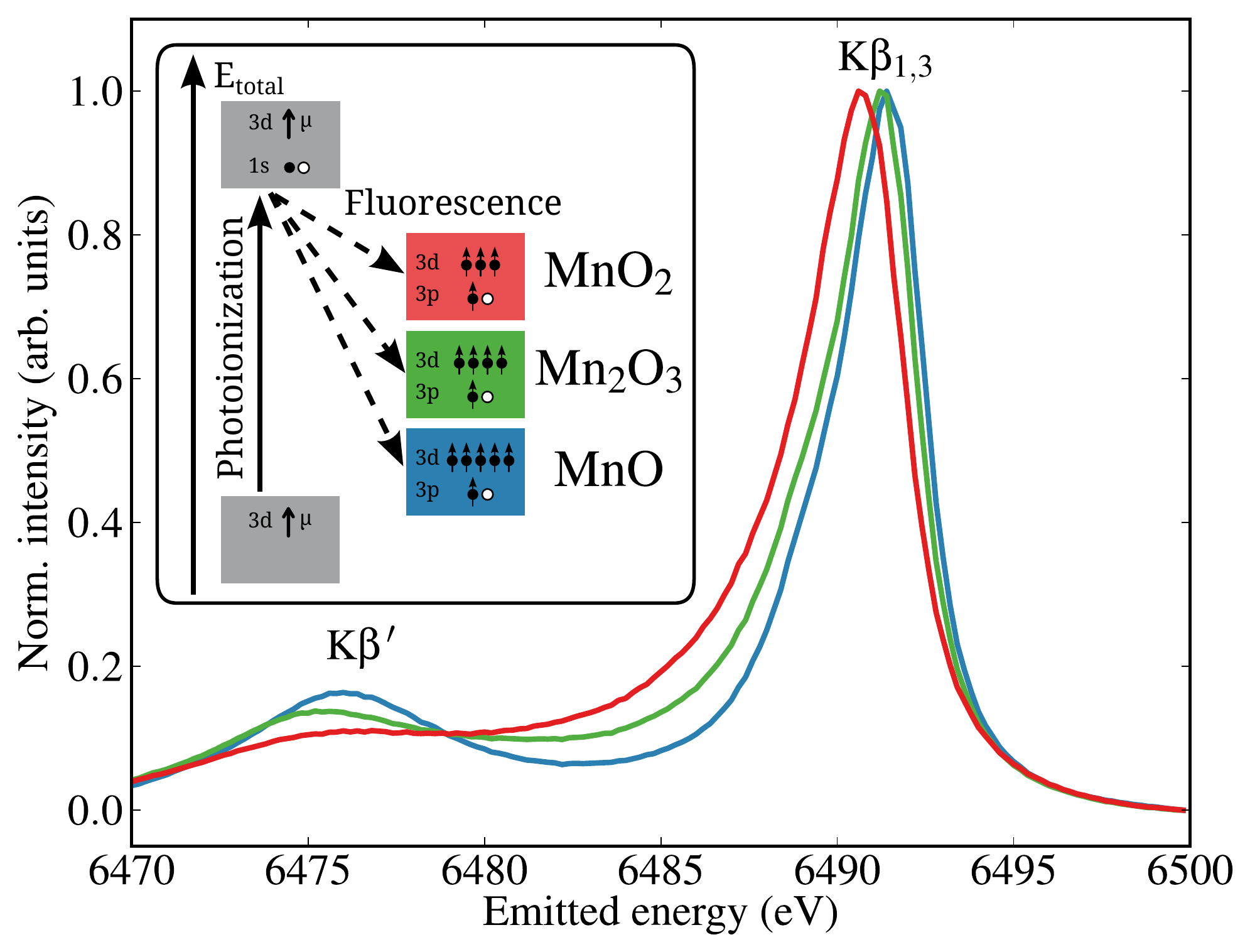}
\caption{Evolution of the \ctc{K$\beta$} spectra for MnO, Mn$_2$O$_3$
  and MnO$_2$. The inset shows this evolution in terms of the total
  energy; for simplicity, only one spin configuration is taken into
  account and paired electrons are omitted.}
\label{fig:kbeta-evolution}
\end{figure}

\begin{figure}[!hbt]
  \includegraphics[width=\MyFigSize]{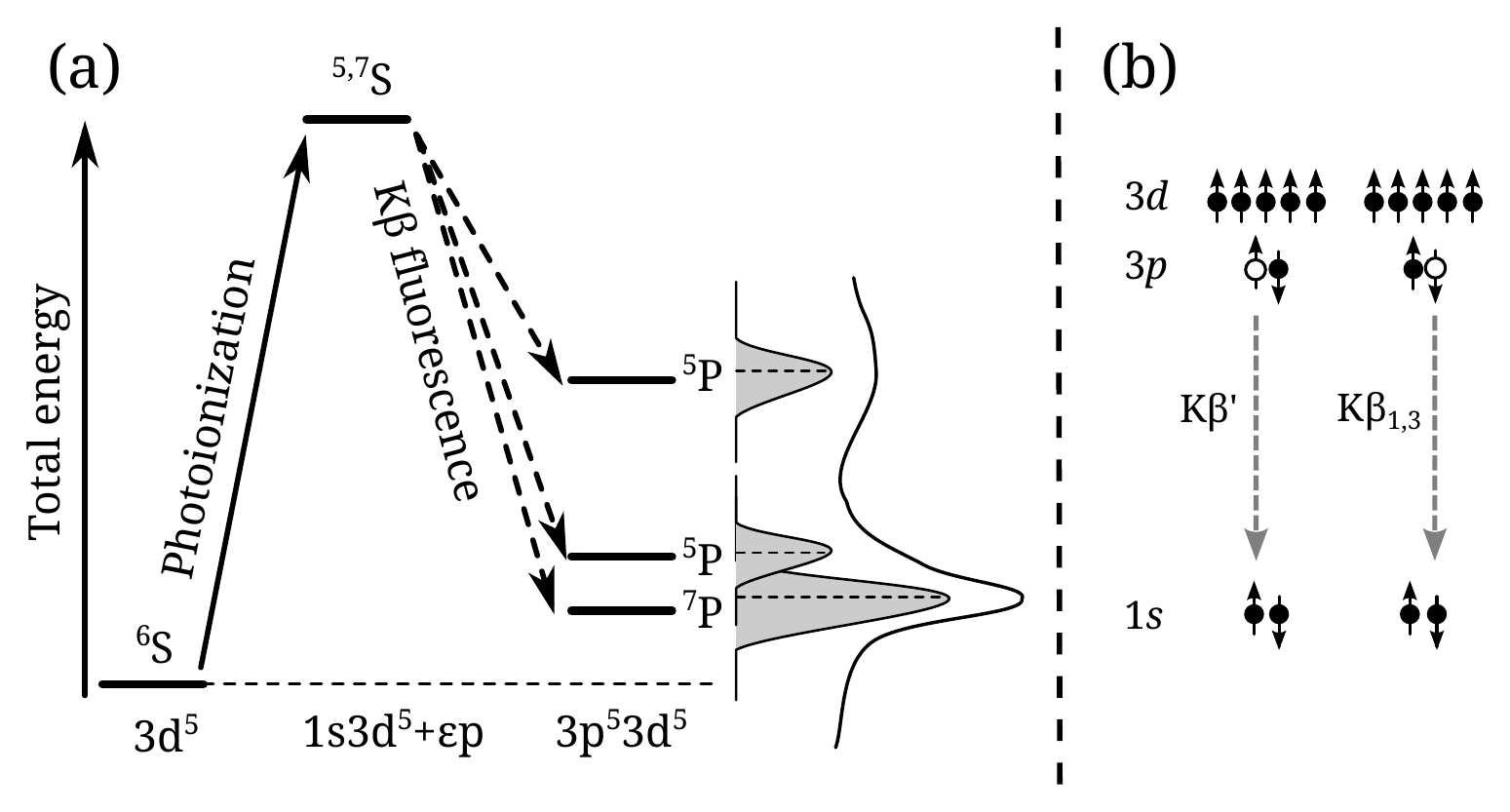}
  \caption{K$\beta$ emission process after photoionization illustrated
    for a 3$d^5$ case. Left panel (a): atomic multiplet theory. Right
    panel (b): simplified one-electron picture (not in energy scale)
    where paired electrons and the spin flipping in K$\beta_{1,3}$ are
    omitted for clarity. A detailed description is given in the text.}
  \label{fig:kbeta-spin}
\end{figure}

%===================================================================%
%
The analysis of the \ctc{K$\beta$} is of particular interest for
magnetic semiconductors because it permits probing (indirectly) the
local magnetic moment brought by the 3$d$ TMs impurities without the
need of demanding sample environment as low temperature and high
magnetic field. In fact, \ctc{K$\beta$} is sensitive to the net local
3$d$ spin moment, independently of its direction. This gives the
possibility to study a magnetic material even in the paramagnetic
state, that is, when the local moments are fluctuating and pointing in
random directions. As shown in figure~\ref{fig:kbeta-evolution} for
three Mn-Oxide powder samples (MnO, Mn$_2$O$_3$, MnO$_2$), the
K$\beta$ main lines evolve with the decreasing nominal spin state
($S$, $S_{\rm MnO}$ = 2.5 $\to$ $S_{\rm MnO_2}$ = 2): the
K$\beta_{1,3}$ shifts toward lower energy and the K$\beta^\prime$
reduces in intensity; this means that the center of mass energy (the
sum of the energies of all final states weighed by their intensities)
does not change between the configurations but the
K$\beta_{1,3}$-K$\beta^\prime$ splitting decreases with decreasing
spin state. This behaviour is understood in a total energy diagram
(inset of figure~\ref{fig:kbeta-evolution}). The intra-atomic exchange
energy between the 3$p$ hole and the 3$d$ levels (sum of the Slater
exchange integrals, $J$) lowers the total energy. As a consequence,
the configurations with parallel spins are lower in energy than the
configurations with paired spins.\\
The K$\beta$ transition involves core levels and multiplet
ligand-field theory is therefore the appropriate framework to discuss
the spectral features
\cite{DeGroot:1994_JPCM,Peng:1994_JACS,Wang:1997_PRB}. In order to
better understand the spin-polarised origin of the K$\beta$ emission
for a 3$d$ TM, we take as example the 3$d^5$ configuration, an atomic
picture and a two-step process \cite{Glatzel:2001_PRB} (as shown in
figure~\ref{fig:kbeta-spin}a). A 3$d^5$ high spin configuration is a
favorable case because of the absence of an orbital angular momentum
in the ground state. Hund's rule dictates that all spins are aligned,
giving the \ts{6}{S}{} ground state spin-orbit term. Photoionization
then excites the system to the 1$s$3$d^5$+$\varepsilon p$
(\ts{5,7}{S}{}) intermediate states. The \ts{7}{S}{} term can only
decay into the \ts{7}{P}{}, while the \ts{5}{S}{} one decays into all
interacting \ts{5}{P}{} states (where symmetry mixing is
important). In consequence, the K$\beta^\prime$ originates almost
100\% from spin-up transitions and the K$\beta_{1,3}$ primarly from
spin-down. This result is exemplified in figure~\ref{fig:kbeta-spin}b
via a one-electron picture of the final state. In
Ref.~\onlinecite{Wang:1997_PRB}, the strong spin selectivity was
demonstrated valid also when the atom is inserted in a crystal
field. In fact, neglecting orbital mixing, the K$\beta$ lines do not
depend on the fine structure in the valence shell (\eg crystal field
splitting) as long a the spin state does not change. On the other
hand, a strong crystal field splitting may result in a low spin
configuration which will change the K$\beta$ line shape
\cite{Badro:2003_S}. These considerations are valid for O$_h$
symmetry. For T$_D$ symmetry, the strong spin polarization is still
present because the crystal field splitting, 10Dq, is simply inverted
between the two symmetries. In addition, the absence of inversion
symmetry (in contrast to O$_h$) results in strong $pd$
mixing. Including orbital mixing in the theoretical description, may
find that the 3$p$3$d$ exchange interaction changes owing to the
mixing and thus the shape of K$\beta$ lines varies. In conclusion, the
spin selectivity is conserved and it is employed to record
spin-selective XAS (\cf \S~\ref{sec:spin-xas}), while the spectral
features change and the methods to take them into account are
discussed in the following.\\
To perform quantitative analysis of the \ctc{K$\beta$}, the first
method is to approximate the energy separation and the intensity ratio
between K$\beta_{1,3}$ and K$\beta^\prime$ by $\Delta E = J(2S+1)$ and
$I^\prime/I_{1,3} = S/(S+1)$ \cite{Tsutsumi:1976_PRB}. This
approximation is found to reproduce fairly well the experimental
results if a peak fitting procedure is employed
\cite{Gamblin:2001_JESRP,TorresDeluigi:2006_CP,Bergmann:1998_JPCB}. On
the other hand, a peak fitting procedure is prone to errors in the
extraction of the peaks positions and arbitrary in the choice of the
number and form of the fitted functions. To overcome the problem of
linking the data analysis to a theoretical approximation, fully
experimental data reduction methods were put in place. The first
attempt was to use the first moment energy of the K$\beta_{1,3}$
\cite{Glatzel:2001_PRB,Messinger:2001_JACS} ($\avg{E}$), defined as
the energy average weighted by the spectrum intensity: $\avg{E} =
\sum_j(E_jI_j) / \sum_j I_j$. Recently, another and more accurate
procedure was proposed \cite{Vanko:2006_JPCB, Vanko:2006_PRB}. It is
based on the integrated absolute difference (IAD) of spectra
\begin{equation}\label{eq:iad}
{\rm IAD}_i = \int_{E_1}^{E_2} \left| \sigma^{\rm XES}_{i}(E) - \sigma^{\rm XES}_{0}(E) \right| dE
\end{equation}
where the XES spectrum $\sigma_0$ is taken as reference (IAD$_{0}$ =
0) and $\sigma_{i}$ is the spectrum for which the IAD value is
determined. Often the IAD values are determined versus a given
parameter within a series (\eg pressure, temperature, concentration,
doping) but the method can be applied to any spectra. It is based on
the differences in the whole spectral range and results in a more
robust procedure, especially when dealing with weak moments. It was
successfully applied to determine the evolution of the local magnetic
moment ($S$) in iron-based superconductors
\cite{Gretarsson:2011_PRB,Chen:2011_PRB,Simonelli:2012_JPCM} or in
strongly correlated oxides
\cite{Lengsdorf:2007_PRB,Sikora:2008_JAP,Herrero-Martin:2010_PRB}.
\begin{figure}[!hbt]
  \includegraphics[width=\MyFigSize]{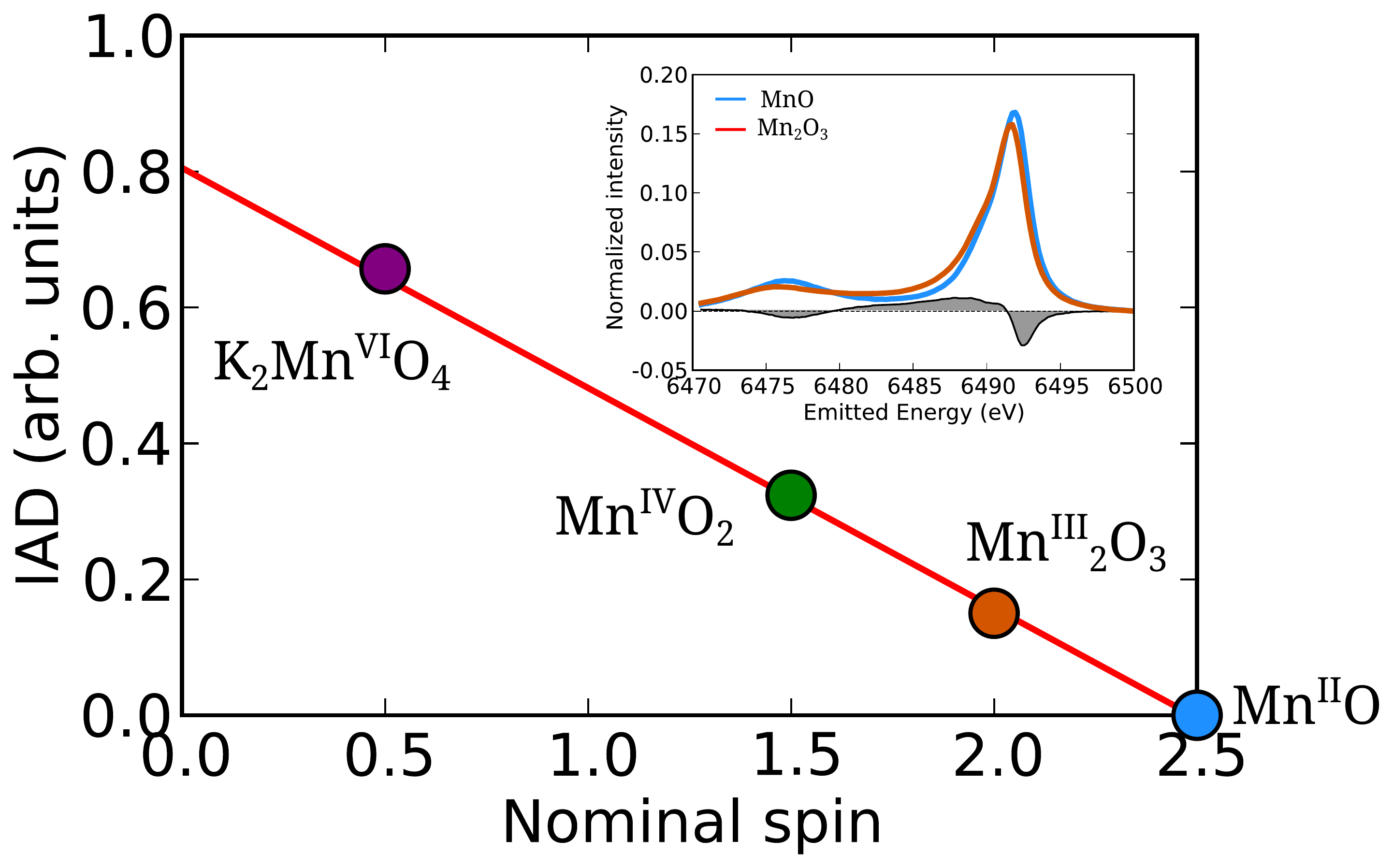}
  \caption{IAD analysis for a series of Mn-Oxides with respect to
    MnO. The inset shows how the IAD is obtained for Mn$_2$O$_3$.}
  \label{fig:IADanalysis}
\end{figure}
To illustrate the method in practice, the IAD analysis for a series of
polycrystalline Mn-Oxides (commercially available powders) is shown in
figure~\ref{fig:IADanalysis}. The IAD values of the normalised spectra
are obtained using MnO as reference (IAD$_{\rm MnO}$ = 0) and are
related to the nominal spin state, assuming a ionic approximation and
a high-spin scenario. Subsequently, a linear fit permits obtaining a
relative calibration that accounts for all possible effects: changes
in oxidation state, bond lengths and angles, site symmetry, energy
shifts during the experiment. By taking into account all these
effects, the error bar on the IAD values is comparable with the size
of the symbols of figure~\ref{fig:IADanalysis}. This makes such analysis
very accurate and reproducible. Once the IAD values are calibrated on
model compounds, it is possible to follow the evolution on real
samples. For DMS/CMS one usually wants to follow the evolution versus
the magnetic dopant concentration or the ratio with shallow impurities
in the case of co-doping.\\
To overcome the crude ionic approximation and to take into account the
covalent bonds in a material (charge transfer), better results are
obtained if the IAD values are compared or calibrated to an effective
local spin moment, $S^{\rm eff}$, defined as
\cite{Limandri:2010_ChemPhys}
\begin{equation}\label{eq:Seff}
S^{\rm eff} = \frac{1}{2} \left( \rho^\uparrow_{A,l} - \rho^\downarrow_{A,l} \right)
\end{equation}
where $\rho^{\uparrow(\downarrow)}_{A,l}$ is the calculated spin
density (charge) on the atom $A$ and projected on the orbital angular
momentum $l$. The projection over $l$ permits having an effective
quantity comparable to the spectroscopic measurement. In fact,
although the charge of an atom in a crystal or molecule is not a good
quantum mechanical observable \cite{Parr:2005_JPCA,Matta:2006_JPCA},
the inner-shell spectroscopist is tempted to assign atomic
properties. Many quantum chemical approaches exist
\cite{Gross:2002_IJQC}, but in DFT the standard methods to perform a
population analysis are those introduced by Mulliken
\cite{Mulliken:1955_JCP}, L\"owdin \cite{Lowdin:1950_JCP} and Bader
\cite{Bader:1991_CR}. In the Mulliken or L\"owdin analysis the charges
are equally divided between two atoms of a bond; this has the
advantage of simplicity. A different approach is followed for the
Bader populations: the electron densities are integrated in a volume
defined by the gradient of the electronic density function. This
scheme usually gives the best results.\\
\begin{figure}[!hbt]
  \includegraphics[width=\MyFigSize]{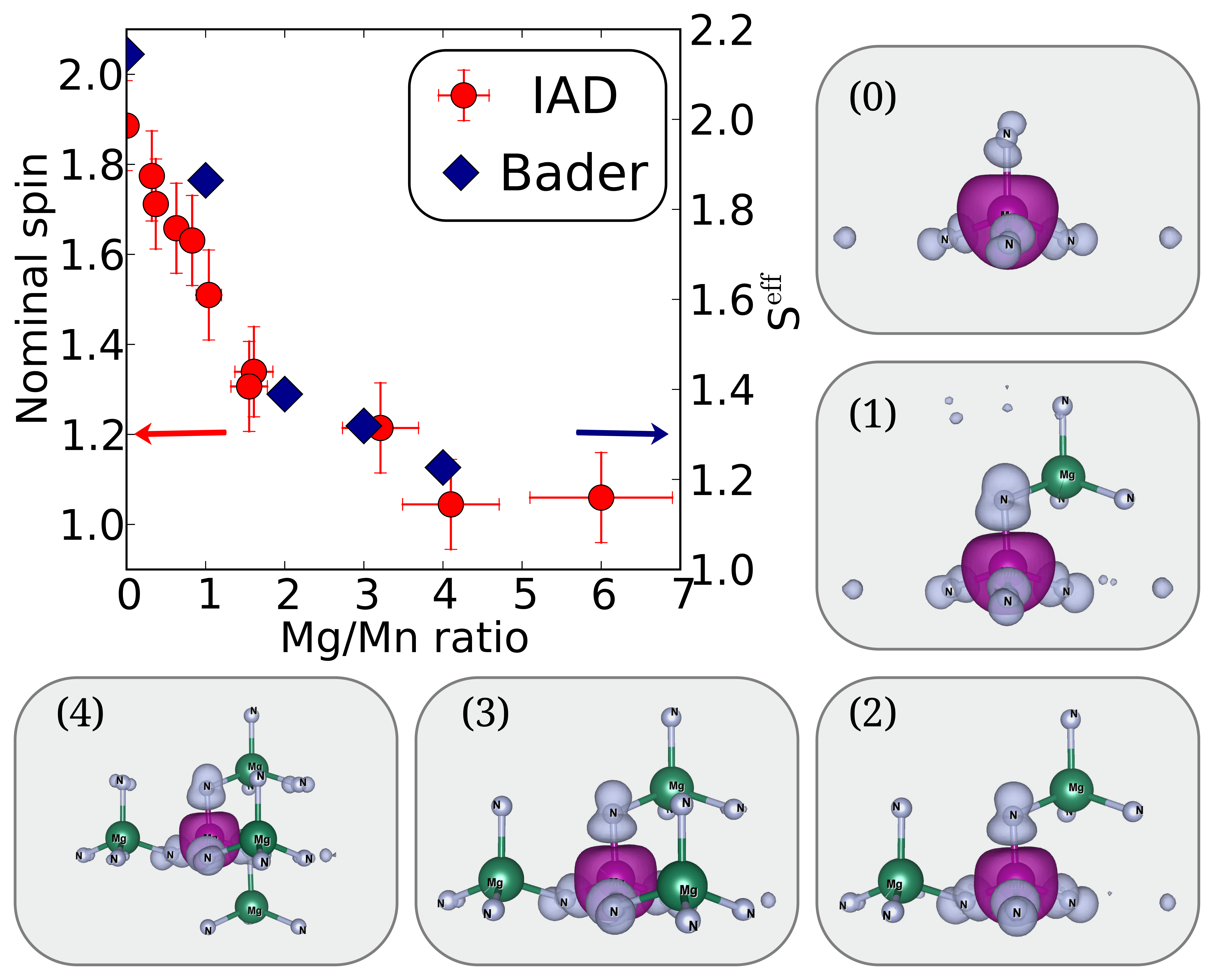}
  \caption{Nominal spin (left, red) extracted by calibrating the IAD
    values via a ionic approximation, compared to S$^{\rm eff}$
    calculated via Bader's population analysis (right, blue). The gray
    panels (0 -- 4) show the calculated polarization on Mn 3$d$ levels
    (violet isosurface) over a ball-and-stick local atomic
    representation of MnN$_4$(MgN$_4$)$_{0 \ldots 4}$ substitutional
    complexes. The surrounding GaN lattice has been removed for
    clarity.}
  \label{fig:gamnn-xes}
\end{figure}
The combination of the IAD analysis with $S^{\rm eff}$ calculated {\em
  ab initio} using DFT has been recently applied in the study of Mn-Mg
substitutional complexes in \ddms{Ga}{Mn}{Mg}{N}
\cite{Devillers:2012_SciRep}. As shown in figure~\ref{fig:gamnn-xes},
the IAD are employed to follow the evolution of the Mn spin state as a
function of the ratio between the Mg and Mn concentration in GaN. By
calibrating the IAD values to the Mn-Oxides reference compounds using
a ionic approximation, it is possible to extract a nominal spin
state. This is then compared to $S^{\rm eff}$ calculated via DFT. Both
evolve in the same way and differ only by a rigid shift ($\approx$ 0.2
in this case). This shift originates from the ionic approximation used
to calibrate the IADs. In fact, the correct procedure to extract a
better absolute measurement of $S$ is to calibrate the IAD via the
Bader analysis performed also on the model compounds. By doing so, one
finds $S_{\rm MnO}$~=~2.2, in perfect agreement with $S^{\rm eff}$
calculated for Mg/Mn~=~0 in figure~\ref{fig:gamnn-xes}. This confirms
that the IAD analysis with an {\em ab initio} $S^{\rm eff}$ is
accurate in following the evolution of the local spin moment.\\
As described for \vtc{K$\beta$}, also in \ctc{K$\beta$} it is possible
to make use of the polarization dependence. For example,
Herrero-Martin {\em et al.} \cite{Herrero-Martin:2010_PRB} studied the
spin distribution in \dms{La}{Sr}{MnO$_{4}$}. They found that
increasing the number of holes (\ie increasing $x$) changes the total
charge (and spin) on Mn very little, but the tetragonal distortion,
that is greatly reduced when going from $x$~=~0 to $x$~=~0.5, causes
an anisotropic spin distribution that also disappears upon hole
doping.
%
%===========================================%
\subsection{Spin-selective XAS}
\label{sec:spin-xas}
%===========================================%
%
Spin-selective XAS was first exploited by H\"am\"al\"ainen and
co-workers \cite{Hamalainen:1992_PRB} and then described via
ligand-field multiplet theory
\cite{Peng:1994_APL,Peng:1994_JACS}. This technique is based on the
strong spin polarization of \ctc{K$\beta$} emission lines (as
previously described in \S~\ref{sec:kbeta}): by collecting a HERFD-XAS
spectrum tuning the spectrometer to the K$\beta_{1,3}$ and
K$\beta^\prime$, it is possible to select, respectively, the
transitions to the spin down and spin up empty density of states (in a
one-electron picture). We underline that the spin-selectivity in this
technique arises only from the K$\beta$ spectrum, that is, the spin
has a local internal reference that does not change in energy for a
change in the direction of the spin moment. With respect to XMCD,
circularly polarised light and an external magnetic field (external
reference) are not required. The link between the two techniques
resides in the energy dependence of the Fano factor
\cite{DeGroot:1995_PRB}.\\
\begin{figure}[!hbt]
  \includegraphics[width=\MyFigSize]{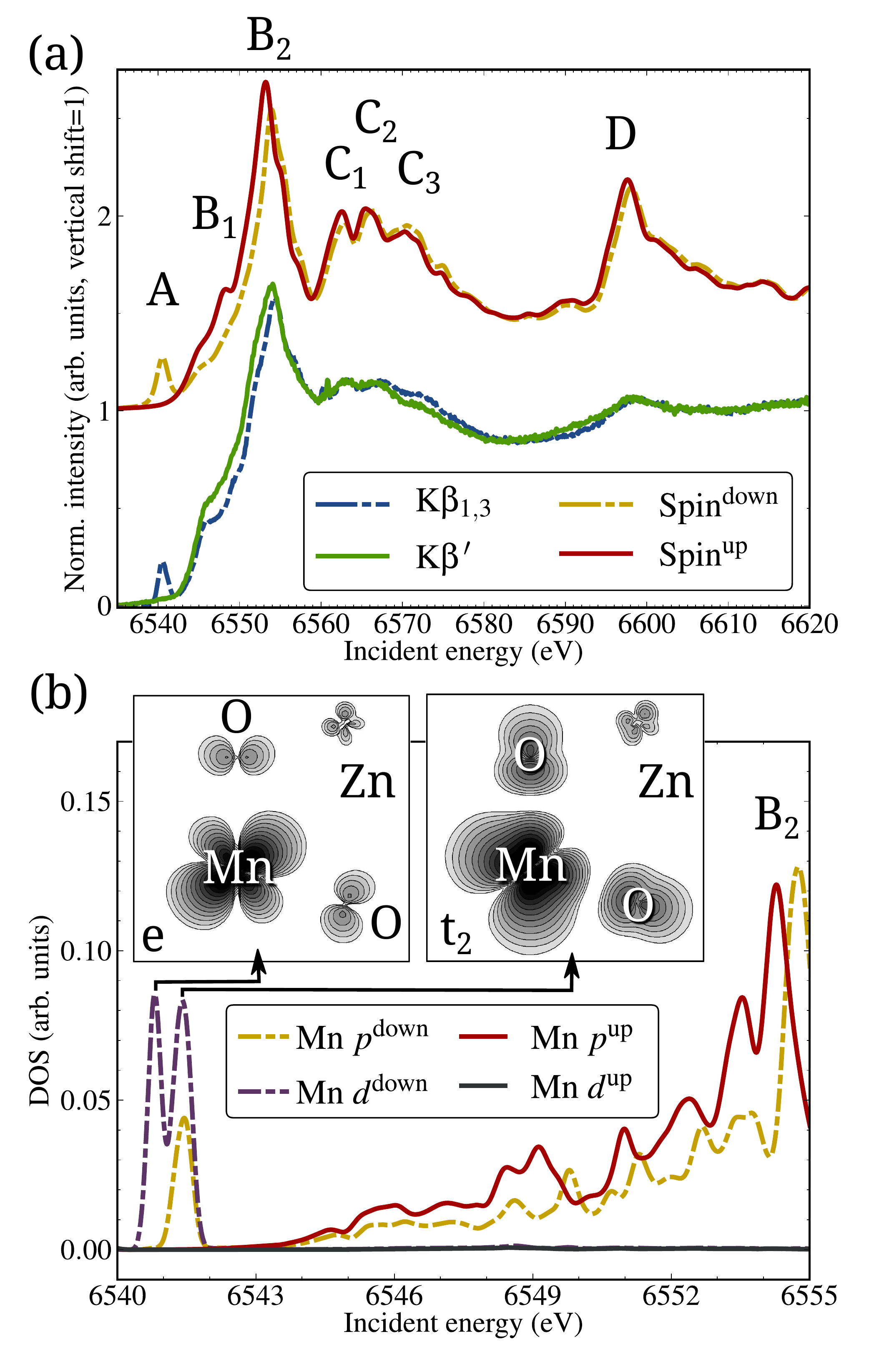}
  \caption{Mn K-edge spin-selective XAS. Top panel (a): experimental
    spectra measured selecting the K$\beta_{1,3}$ and K$\beta^\prime$
    emission channels (bottom lines), compared to the spin-polarised
    FLAPW simulations for a Mn$_{\rm Zn}$ substitutional defect in
    ZnO, respectively, for the spin down and spin up densities (top
    lines). Bottom panel (b): magnification in the pre-edge region of
    the spin-polarised Mn$_{\rm Zn}$ density of states projected to
    $p$ and $d$ orbital angular momenta. The insets show the electron
    density distribution (log scale) in the $xz$ plane for the $e$ and
    $t_{2}$ levels (splitted by the crystal field) of Mn$_{\rm
      Zn}$. Adapted figure~4 from Ref.~\onlinecite{Guda:2013_JAAS} with
    permission from The Royal Society of Chemistry.}
  \label{fig:spin-XAS}
\end{figure}
An example of application of this technique to the characterization of
magnetic semiconductors was reported recently
\cite{Guda:2013_JAAS}. The Mn K-edge HERFD-XAS spectra of
ZnO/\dms{Zn}{Mn}{O} core/shell nano-wires were measured at
K$\beta_{1,3}$ and K$\beta^\prime$, then compared to {\em ab initio}
DFT calculations using a FLAPW approximation. As shown in
figure~\ref{fig:spin-XAS}, the spectral features A, B$_{1,2}$,
C$_{1\dots3}$ and D are reproduced by the theory (panel {\em a}), in
agreement between the two spin-polarizations. This agreement was
obtained using a Mn defect substitutional of Zn (Mn$_{\rm Zn}$) in a
ZnO relaxed supercell. Of particular interest for DMS is the pre-edge
region where the electronic structure of the spin-polarised Mn$_{\rm
  Zn}$ impurity level can be studied in detail. In the case of
zincblende and wurtzite DMS, the local symmetry around the cation is
tetrahedral (T$_D$) in first approximation (we do not take into
account the Jahn-Teller effect \cite{Virot:2011_JPCM}), this means
that the TM $d$ states are splitted by the crystal field into $e$
(doublet) and $t_2$ (triplet) levels. As reported in the panel {\em b}
of figure~\ref{fig:spin-XAS}, they are fully spin polarised. In
addition, because of the $t_2$ symmetry, they can partially couple
with the $p$ bands \cite{Westre:1997_JACS}. Electric dipole
transitions to $t_2$ dominate over quadrupole ones to $e$
\cite{Hansmann:2012_PRB}, this explains why the spin-selective XAS
technique is extremelly sensitive in the pre-edge region at the fine
details of the electronic structure of these materials.\\
%
%======================================================%
\section{RIXS and magnetic circular dichroism}
\label{sec:rixs-mcd}
%======================================================%
%
%
\begin{figure*}[!hbt]
  \includegraphics[width=\MyFigSizeStar]{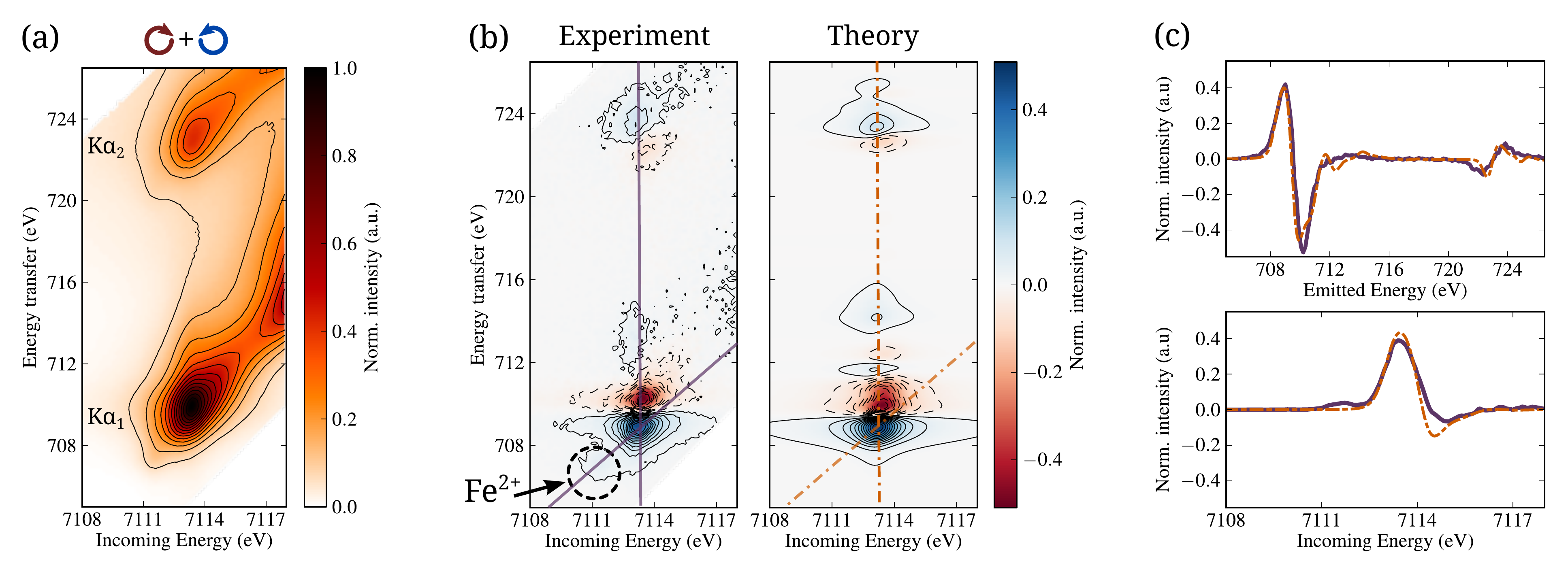}
  \caption{RIXS-MCD study of Fe$_3$O$_4$ (magnetite) powder. Left
    panel (a): experimental total absorption, that is, right (\cpolr)
    plus left (\cpoll) circular polarization. Middle panel (b):
    experimental and simulated circular dichroism, \cpolr-\cpoll. The
    simulation is only for Fe$^{3+}_{\rm T_D}$, while the experimental
    data show an additional feature attributed to Fe$^{2+}$
    (highlighted) - \cf main text. Right panel (c): CIE (vertical) and
    CEE (diagonal) line cuts, respectively, at 7113.8 eV and 6404.8 eV
    (K$\alpha_1$ maximum). Adapted Figs.~1~and~2 from
    Ref.~\onlinecite{Sikora:2010_PRL}, with permission from the
    American Physical Society and data from the authors.}
  \label{fig:rixs-mcd}
\end{figure*}
RIXS can be coupled with magnetic circular dichroism (RIXS-MCD) using
circularly polarised x-rays and an external magnetic field applied to
the sample. In selected cases, RIXS-MCD permits combining the
benefits of hard and soft x-rays XMCD by selecting specific final
states. The idea is to make use of hard x-rays and reach a spin-orbit
split final state indirectly via an intermediate state (\cf total
energy diagram in figure~\ref{fig:Mn_energy_dia}), where the dichroism
arises from the coupling of the magnetic moment of the absorbing atom
with the circularly polarised light. The first application of RIXS-MCD
was at the L$_{2,3}$ pre-edges of RE elements, where the quadrupole
transition channel \cite{Krisch:1995_PRL} (2$p$~$\to$~4$f$), was used
to probe with hard x-rays and via a second-order process the same
final states obtained with direct dipole transitions in M$_{4,5}$
lines XMCD
\cite{Caliebe:1996_JAP,Krisch:1996_PRB,Iwazumi:1997_PRB,Fukui:2001_JSR,Fukui:2004_JPSJ}. Recently,
this approach was successfully applied also to the K pre-edges of TM
elements \cite{Sikora:2010_PRL}, extending the possibilities of K-edge
XMCD. For TMs, the required intermediate state can be excited either
via quadrupole transitions (1$s$~$\to$~3$d$) or dipole to mixed
3$d$-4$p$ states. This means that RIXS-MCD cannot be applied to
metals, but to TMs-doped semiconductors and insulators or bulk TMs
oxides and nitrides.\\
In figure~\ref{fig:rixs-mcd} we show the application of RIXS-MCD to
magnetite (Fe$_3$O$_4$) as reported in
Ref.~\onlinecite{Sikora:2010_PRL}. Magnetite is a ferrimagnetic
inverse spinel with a Curie temperature of 860~K, high spin
polarization at room temperature, good magnetostriction and showing a
metal-insulator transition at about 120~K \cite{Verwey:1939_Nat}
(Verwey transition). These properties make the material very appealing
for spintronics heterostructures. It is a challenging material also
from the characterization point of view (\cf
Refs.~\onlinecite{Senn:2012_Nat,Bengtson:2013_PRB} plus references
therein). It is commonly accepted that Fe$_3$O$_4$ has two differently
coordinated and antiferromagnetically coupled sublattices (T$_D$ and
O$_h$), with mixed valences on the O$_h$ site (Fe$^{2+}$ and
Fe$^{3+}$) and with Fe$^{2+}$ mainly responsible for the resulting
magnetic moment. In synthesis, its formula can be written as
Fe$^{3+}_{\rm T_D}$Fe$^{2+,3+}_{\rm O_h}$O$_4$. The total absorption
(figure~\ref{fig:rixs-mcd}a) shows the pre-edge region resolved in the
K$\alpha_{1,2}$ spin-orbit split emission lines, that is, 1$s$2$p$
RIXS. The detailed structure of the Fe pre-edge features was
extensively described for XAS \cite{Westre:1997_JACS} and RIXS
\cite{DeGroot:2005_JPCB}. We can summarise, in a nutshell, that the
features arising at $\approx$~7114~eV incident energy and
$\approx$~710~eV energy transfer are from Fe$^{3+}$, while those at
$\approx$~7112~eV and $\approx$~707~eV from Fe$^{2+}$. This shows as a
broadening in the diagonal direction in the RIXS plane. The effect is
better visible on the K$\alpha_1$ line (due to a better
signal-to-noise ratio).\\
In figure~\ref{fig:rixs-mcd}b the circular dichroism (right minus left
circular polarization) is shown for the experiment and the theory. The
simulated 1$s$2$p$ RIXS-MCD plane is based on ligand-field multiplet
calculations for Fe$^{3+}_{\rm T_D}$ only. As demonstrated in
Ref.~\onlinecite{Sikora:2010_PRL}, the strong dichroism originates in
part from the sharpening effect (as described in
\S~\ref{sec:rixs-plane}) but mainly from the 3$d$ spin-orbit
interaction in the intermediate state and the 2$p$-3$d$ Coulomb
repulsion combined with the 2$p$ spin-orbit interaction in the final
state. The main spectral features are reproduced by the calculation,
confirming that the Fe$^{3+}_{\rm T_D}$ site is dominating the
measured signal. By taking a vertical cut along the energy transfer
(figure~\ref{fig:rixs-mcd}c, top), the resulting line scan is comparable
with a L$_{2,3}$ XMCD spectrum, both in the sign - plus/minus
(minus/plus) for K$\alpha_1$ (K$\alpha_2$) - and the amplitude. The
enhancement in amplitude with respect to a conventional K-edge XMCD is
one of the advantages of RIXS-MCD. In addition, the RIXS-MCD plane
shows an extra feature in the region ascribed to Fe$^{2+}$ (\cf
figure~\ref{fig:rixs-mcd}b). This feature is not reproduced when only
Fe$^{3+}_{\rm T_D}$ site is taken into account and its weak intensity
permits interpreting it as originating from Fe$^{2+}_{\rm O_h}$ mainly
\cite{Sikora:2010_PRL}. A diagonal cut at the K$\alpha_1$ maximum also
shows it (figure~\ref{fig:rixs-mcd}c, bottom).\\
This demonstrates that is possible to use RIXS-MCD both as element-
and valence-/site-selective magnetometer by means of field dependent
measurements \cite{Sikora:2012_JAP}. In magnetic semiconductors, this
technique would permit disentangling the extrinsic magnetism coming
from metallic precipitates from the intrinsic one.\\
%
%======================================================%
\section{Conclusions and future developments}
\label{sec:final}
%======================================================%
%
In this review we have presented the basic elements of XES and RIXS
spectroscopies, with a focus on the characterization of magnetic
semiconductors and employing hard x-rays. The theoretical background
(Kramers-Heisenberg equation), the required experimental setup and the
approaches to the calculations are the building blocks for a practical
introduction to the field. With respect to doped semiconductors, XES
and RIXS play a crucial role in studying the local electronic
structure around the Fermi level. By using the \vtc{K$\beta$} is
possible to directly probe the ligand states in the valence band,
while the \ctc{K$\beta$} is a sensitive tool of the local spin angular
momentum, via intra-atomic exchange interaction. We have also shown
that with RIXS (or spin-selective XAS) is then possible to complement
the electronic structure picture by probing unoccupied states
(conduction band) with spin sensitivity. Finally, RIXS-MCD permits
extending the element-selective magnetometry (XMCD) by gaining in
signal intensity plus in site and valence selectivity.\\
Supported by the fast evolution in the theoretical tools and the
development of new instrumentation, the users community in this field
is growing. This gained momentum permits not only better analysis of
current data, but also prediction and realization of new challenging
experiments. In particular, the materials scientist working with
strongly correlated materials will benefit of a photon-in/photon-out
spectroscopy in the hard x-ray spectrum. In fact, this technique will
permit characterizing, at the atomic level, devices in operating
conditions (\eg a spin field-effect transistor with applied gate
voltage), with the possibility to perform direct tomography
\cite{Huotari:2011_NMat}. Magneto-optical devices can be characterised
in a laser pump and x-ray probe configuration to study fast spin
dynamics as spin-orbit interaction \cite{Boeglin:2010_Nat} or spin
state transitions \cite{Vanko:2012_JESRP}.\\
RIXS experiments require the high brilliance of third generation
synchrotron radiation sources or even x-ray free electron lasers, that
is, high photon flux with small divergence and small energy
bandwidth. However, XES is a powerful tool which can be accessible
also outside large scale facilities. In fact, the advantage of XES is
that it can be performed with a pink beam. This permits adapting an
XES instrument on any synchrotron radiation beamline (\eg standard
XAS, x-ray diffraction, imaging) or on a laboratory x-ray tube. For
example, one can take the case of measuring Mn \ctc{K$\beta$} on a
Ga$_{0.97}$Mn$_{0.03}$N thin film (dilute material in a strong
absorbing matrix). By a simple comparison only on the incoming photon
flux, assuming 10$^{9}$~ph/s (\eg from a rotating anode x-ray tube)
one would get $\approx$~10 counts/s on the Mn \ctc{K$\beta$}
maximum. Considering the very low background of a point-to-point
spectrometer, a spectrum with a reasonable signal to noise ratio is
obtained in one day of measurements.
%
%======================================================%
\ifnum\MyClass=1
\ack %IOP class
\fi
\ifnum\MyClass=0
\begin{acknowledgments}
\fi
\ifnum\MyClass=2
\section*{Acknowledgments}
\fi
We gratefully acknowledge the European Synchrotron Radiation Facility
for providing synchrotron radiation via {\em in house research}
projects. M.R. would like to thank M.~Sikora, E.~Gallo and
N.~Gonzalez~Szwacki for fruitiful discussions.
\ifnum\MyClass=0
\end{acknowledgments}
\fi
%======================================================%
%<revtex4-1>
\ifnum\MyClass=0
\bibliographystyle{apsrev4-1}
\fi
%</revtex4-1>
%<iopart>
\ifnum\MyClass=1
\References
\bibliographystyle{unsrt}
\fi
%</myiopart>
\ifnum\MyClass=2
\bibliographystyle{iopart-num}
\fi
%</myiopart>

% Bitex database
\bibliography{XESrev4SST-v10arxiv.bib}

% Bibliography

%%% END %%%
\end{document}